\documentclass[12pt,a4]{article}
\usepackage{amsthm,amsmath,latexsym,amssymb,amsfonts,amscd}
\usepackage{graphics,fancyhdr,array,stmaryrd,euscript}
\usepackage{slashed,tikz}
\numberwithin{equation}{section}

\usepackage{hyperref}

\usepackage[numbers,sort&compress]{natbib}
\usepackage[nottoc]{tocbibind}
\usepackage{xcolor}

\newcommand{\pl}{\partial}
\newcommand{\be}{\begin{equation}}
\newcommand{\ee}{\end{equation}}
\newcommand{\bea}{\setlength\arraycolsep{2pt} \begin{eqnarray}}
\newcommand{\eea}{\end{eqnarray}}
\newcommand{\eq}[1]{(\ref{#1})}
\newcommand{\w}[1]{\\[0.#1cm]}

\newcommand{\mm}{{\ensuremath{\underline{m}}}}
\newcommand{\nn}{{\ensuremath{\underline{n}}}}

\newcommand{\aI}{{\ensuremath{\mathcal{I}}}}

\newcommand{\gad}{{\dot{\alpha}}}
\newcommand{\gbd}{{\dot{\beta}}}
\newcommand{\gdd}{{\dot{\gamma}}}
\newcommand{\gdl}{{\dot{\delta}}}

\newcommand{\gmd}{{\dot{\mu}}}
\newcommand{\gnd}{{\dot{\nu}}}

\newcommand{\ga}{\alpha}
\newcommand{\gb}{\beta}
\newcommand{\gc}{\gamma}
\newcommand{\gd}{\delta}

\newcommand{\fd}[1]{{}^{\vphantom{#1}}_{#1}}
\newcommand{\fud}[2]{{}^{#1}{}_{#2}\,}
\newcommand{\fdu}[2]{{}_{#1}{}^{#2}\,}

\newcommand{\brz}{{{\bar{z}}}}

\newcommand{\brxi}{{{\bar{\xi}}}}
\newcommand{\brmu}{{{\bar{\mu}}}}

\newcommand{\besubeqs}{\begin{subequations}}
\newcommand{\esubeqs}{\end{subequations}}

\newcommand{\VVV}{{\boldsymbol{V}}}
\newcommand{\VVVf}{{\boldsymbol{\mathbb{V}}}}
\newcommand{\VVVb}{{\boldsymbol{\bar{V}}}}
\newcommand{\VVVh}{{\boldsymbol{\hat{V}}}}
\newcommand{\VV}{{\boldsymbol{\tilde V}}}%this is for ads vertices

\newcommand{\tanV}{\mathrm{F}}
\newcommand{\Fron}{{\Phi}}

\newcommand{\Pib}{{{\bar{\Pi}}}}
\newcommand{\xisc}[2]{(\xi_{#1} \xi_{#2})}
\newcommand{\xiscb}[2]{(\brxi_{#1} \brxi_{#2})}

\newcommand{\JJ}{{\mathbb{J}}}

\newcommand{\rmx}{{\mathrm{x}}}
\DeclareMathOperator{\sign}{sign}
\newcommand{\vol}{{\text{vol}\,}}
\newcommand{\Tr}{\mathrm{Tr}}

\begin{document}
%%%%%%%%%%%%%%%%%%%%%%%%%%%%%%%%%%%%%%%%%%%%%%%%%%%%%%%%%%%%%
\pagenumbering{gobble}
\hfill
\vskip 0.01\textheight
\begin{center}
{\Large\bfseries 
On (spinor)-helicity and bosonization in $\boldsymbol{AdS_4/CFT_3}$}

\vskip 0.03\textheight
\renewcommand{\thefootnote}{\fnsymbol{footnote}}
Evgeny \textsc{Skvortsov}\footnote{Research Associate of the Fund for Scientific Research -- FNRS, Belgium}${}^{a,b}$ \& Yihao \textsc{Yin}${}^{c,d}$
\renewcommand{\thefootnote}{\arabic{footnote}}
\vskip 0.03\textheight

{\em ${}^{a}$ Service de Physique de l'Univers, Champs et Gravitation, \\ Universit\'e de Mons, 20 place du Parc, 7000 Mons, 
Belgium}\\
\vspace*{5pt}
{\em ${}^{b}$ Lebedev Institute of Physics, \\
Leninsky ave. 53, 119991 Moscow, Russia}\\
\vspace*{5pt}
{\em ${}^{c}$College of Physics, Nanjing University of Aeronautics and Astronautics, \\
Nanjing 211106, China}\\
{\em ${}^{d}$Key Laboratory of Aerospace Information Materials and Physics (NUAA), MIIT,\\
Nanjing 211106, China}\\

\vskip 0.05\textheight

\begin{abstract}
Helicity is a useful concept both for AdS${}_4$ and CFT${}_3$ studies. 
We work out the complete AdS${}_4$/CFT${}_3$ dictionary for spinning fields/operators in the spinor-helicity base that allows one to scalarize any $n$-point contact vertex. AdS${}_4$-vertices encode correlation functions of conserved currents, stress-tensor and, more generally, higher spin currents in a simple way. We work out the dictionary for Yang-Mills- and gravity-type theories with higher derivative corrections as well as some higher spin examples and exemplify the relation to the three-dimensional bosonization duality. The bosonization can be understood as a simple surgery: vertices/correlators are built via an EM-duality transformation by sewing together (anti)-Chiral higher spin gravities, to whose existence the three-dimensional bosonization duality can be attributed (up to the proof of uniqueness).
\end{abstract}

\end{center}
\newpage
\tableofcontents
\newpage
%%%%%%%%%%%%%%%%%%%%%%%%%%%%%%%%%%%%%%%%%%%%%%%%%%%%%%%%%%%%%
\section{Introduction}
%%%%%%%%%%%%%%%%%%%%%%%%%%%%%%%%%%%%%%%%%%%%%%%%%%%%%%%%%%%%%
\pagenumbering{arabic}
\setcounter{page}{1}

Spinor-helicity formalism, see e.g. \cite{Elvang:2013cua, Henn:2014yza},  is a golden standard to write down the on-shell amplitudes of massless particles with spin, the earliest illustration of its efficiency having been the Parke-Taylor formula \cite{Parke:1986gb}. The effectiveness of the spinor-helicity approach is also linked to twistor space techniques \cite{Witten:2003nn,Roiban:2004ka,Bena:2004ry,Cachazo:2004zb,Adamo:2011cb,Adamo:2016ple}. Spinor-helicity formalism is closely related to the light-cone approach, the latter providing an off-shell extension of the former \cite{Ponomarev:2016cwi}.  Therefore, it is natural to extend the spinor-helicity language to the curved setting, two most important cases being dS${}_4$ and AdS${}_4$. While the two space-times are close relatives, specific applications require slightly different approaches, see e.g. \cite{Maldacena:2011nz,David:2019mos,Albrychiewicz:2021ndv} for dS${}_4$ and e.g. \cite{Nagaraj:2018nxq,Nagaraj:2019zmk,Nagaraj:2020sji} for AdS${}_4$. We will develop further the AdS${}_4$-version of \cite{Didenko:2012tv}, see also \cite{Colombo:2012jx,Didenko:2013bj,Bonezzi:2017vha,Sezgin:2017jgm}, which is 
applicable to dS${}_4$ as well. Our main motivation is to explore three-dimensional bosonization duality \cite{Giombi:2011kc, Maldacena:2012sf, Aharony:2012nh,Aharony:2015mjs,Karch:2016sxi,Seiberg:2016gmd}.

Ideally any reasonable CFT should be dual to a theory of quantum gravity in AdS. Here `being dual' depends on whether or not both sides of the duality have a non-perturbative description or we have to restrict ourselves to perturbation theory and usually it is the latter. Therefore, an important question on the bulk side is: does the corresponding gravity dual admit a tractable semi-classical expansion in some limit. Usually it is not, and the CFT has to obey certain conditions for it to have a well-defined semi-classical local bulk dual of field theory type, see e.g. \cite{Heemskerk:2009pn}. \textit{Semi-classical} is equivalent to having some sort of large-$N$ expansion. \textit{Field theory type} can be phrased as having the large gap in the spectrum with finitely many fields below the gap, i.e. in the first approximation the theory has finitely many fields. \textit{Locality} is more subtle, e.g. one can easily write down nontrivial $\nabla^k \phi^4$-type interactions for arbitrarily large $k$.\footnote{If the stringy picture is available, then the strings' length has to be much smaller then AdS radius. } One idea \cite{Maldacena:2015iua} is that bulk locality implies certain additional singularities of CFT correlators.

It would be a great advantage to have simple toy models of quantum gravity, the requirements for them being to have the graviton in the spectrum\footnote{Some CFT's do not have stress-tensor, e.g. long-range Ising model, and even this requirement can be dropped. In this case one can consider all the usual renormalizable theories in (A)dS${}_4$, e.g. $\phi^4$ \cite{Bertan:2018khc,Bertan:2018afl,Heckelbacher:2020nue,Heckelbacher:2022fbx,Heckelbacher:2022hbq}.} and to be finite/renormalizable. Of course, these features cannot come for free and one has to sacrifice something else, e.g. unitarity. Putting unitarity aside we have a few rather simple theories: (i) conformal $\mathcal{N}=4$ supergravity\footnote{It is a matter of opinion, of course, whether to call these theories simple given the complexity of the action, which is, nevertheless known explicitly thanks to the remarkable efforts in \cite{Butter:2016mtk,Butter:2019edc}. } coupled to four multiplets of $\mathcal{N}=4$ SYM \cite{Fradkin:1985am}; (ii) various self-dual theories ranging from self-dual (super) Yang-Mills theory (SDYM) to self-dual (super)gravities (SDGR) and Chiral higher spin gravity \cite{Metsaev:1991mt,Metsaev:1991nb,Ponomarev:2016lrm,Skvortsov:2018jea,Skvortsov:2020wtf,Skvortsov:2022syz,Sharapov:2022faa} that admits an extension to (A)dS${}_4$ \cite{Metsaev:2018xip,Skvortsov:2018uru,Krasnov:2021nsq,Sharapov:2022awp}. Self-dual theories have a number of attractive properties, e.g. integrability, finiteness, relation to twistors and instantons. Self-dual theories have vanishing tree level amplitudes in flat space, but this is not the case in curved backgrounds, e.g. in AdS${}_4$. Since self-dual theories and Chiral higher spin gravity can be thought of as consistent truncations of their unitary completions, all physical observables in these theories are unitary in the sense of being parts of the unitary ones. 

One more application of Chiral higher spin gravity is to three-dimensional bosonization duality \cite{Skvortsov:2018uru,Sharapov:2022awp}. Its existence proves that Chern-Simons vector models have two closed subsectors of the correlation functions. The very existence of a one-parameter family of CFT's connecting various free/critical vector models via Chern-Simons extensions thereof can be attributed \cite{Skvortsov:2018uru} to existence of Chiral higher spin gravity. This idea is constructive and leads to explicit formulas for the correlators, see \cite{Skvortsov:2018uru} for the complete $3$-point correlators, which gives an independent proof of the results based on slightly-broken higher spin symmetry \cite{Maldacena:2012sf} and extends the results of \cite{Giombi:2016zwa}. The main idea of \cite{Skvortsov:2018uru} is to use the helicity decomposition of bulk interactions/CFT correlators. As such it does not even require any AdS${}_4$-realization and can be phrased on the CFT side. We elaborate more on this in section \ref{sec:Bosonization} up to four-point correlators as well as on the relation between the bulk spinor-helicity formalism and bosonization, see also section \ref{subsec:HS}.

To deal with theories in the AdS${}_4$/CFT${}_3$-context it is advantageous to have a fine-grained decomposition of usual interactions/correlators with respect to their helicity structure, which is useful on its own. Helicity is a special feature of massless fields in four dimensions: all massless fields with nonzero spin have two physical degrees of freedom, which greatly simplifies the structure of amplitudes and is at the core of many important developments starting from spinor-helicity. When properly treated massless fields have the same two degrees of freedom on curved backgrounds, in particular, on AdS${}_4$ or dS${}_4$. The latter should make it possible to extend the efficient flat space techniques to, at least, constant curvature spaces. This idea has recently been put forward in a number of papers, see e.g. \cite{Farrow:2018yni,Armstrong:2020woi,Cheung:2022pdk,Herderschee:2022ntr}. 

Once the concept of helicity is extended to AdS${}_4$, nothing prevents us from propagating it further to CFT${}_3$ \cite{Skvortsov:2018uru}, see also \cite{Caron-Huot:2021kjy,Jain:2021vrv,Jain:2022ujj} for very recent studies in the same vein. Not surprisingly, the helicity decomposition of correlation functions provides a more fine-grained split that makes a number of useful properties manifest. As a result, one gets a one-to-one correspondence between $4d$ massless amplitudes in flat space, vertices/amplitudes in (A)dS$_4$ and CFT${}_3$ structures 
\begin{align*}
    \text{\fbox{\parbox{2.8cm}{\centering $4d$ Minkowski \\Amplitudes}}} \Longleftrightarrow \text{\fbox{\parbox{2.6cm}{\centering  AdS${}_4$ \\  Amplitudes}}} \Longleftrightarrow \text{\fbox{\parbox{2.6cm}{\centering  CFT${}_3$ \\ Correlators}}} 
\end{align*}
This gives a bit more of conformally-invariant structures \cite{Skvortsov:2018uru} as compared to the canonical covariant approaches \cite{Giombi:2011rz}, e.g. stress-tensor three-point function can now be split into $+++$, $++-$, $--+$ and $---$ pieces whereas one usually observes only a sum of $++-$ and $--+$, which is similar to the spinor-helicity decomposition of the Einstein-Hilbert cubic amplitude. 

With the help of the fine-grained helicity decomposition of vertices/correlators the three-dimensional bosonization duality can be understood \cite{Skvortsov:2018uru} as a simple surgery: the helicity decomposition of the correlators/vertices is followed by a simple phase, EM-duality, rotation, after which they are sewed back to correlation functions of a unitary, but parity-violating (in general) CFT, where the key point is that (anti)-chiral subsectors are rigid and closed. An interpretation on the AdS${}_4$-side is exactly the same with correlators replaced by the vertices. Helicity decomposition in \cite{Skvortsov:2018uru} was done within the light-cone approach in AdS${}_4$ \cite{Metsaev:2018xip}, which makes the helicity explicit from the very beginning, but at the cost of sacrificing manifest Lorentz invariance. One of the results of the present paper is to construct a basis of three-point amplitudes and correlators that exhibits the helicity decomposition while maintaining Lorentz invariance. We have to admit, however, that the decomposition is not perfect: in many cases the amplitudes have mixed structure. Nevertheless, this suffices for applications to the bosonization duality. 

After a brief review of an efficient approach \cite{Giombi:2011rz} to CFT${}_3$ correlators in section \ref{sec:review} we proceed to boundary-to-bulk propagators and spinor-helicity formalism of \cite{Didenko:2013bj}. In section \ref{sec:amplitudes} we discuss the general structure of $3$-point amplitudes. To give examples we elaborate on a number of usual suspects in section \ref{sec:correlators}: Yang-Mills theory with higher derivative corrections; gravity with higher derivative corrections. As is clear from the previous studies, e.g. \cite{Maldacena:2011nz}, pure Einstein-Hilbert action does not lead to a three-point function of the stress-tensor in any free CFT, instead it gives a linear combination of the free boson and free fermion contributions. We also provide a higher spin example in section \ref{sec:correlators} and discuss the general properties of the AdS/CFT dictionary. Applications to the bosonization duality are in section \ref{sec:Bosonization}. There is a great number of technical appendices, where we explain our notation, collect important, yet cumbersome, formulas and explain how to scalarize any $n$-point contact interaction vertex in AdS${}_4$.

%%%%%%%%%%%%%%%%%%%%%%%%%%%%%%%%%%%%%%%%%%%%%%%%%%%%%%%%%%%%%%%%%%%%%%%
\section{AdS/CFT Dictionary} 
\label{sec:review}
%%%%%%%%%%%%%%%%%%%%%%%%%%%%%%%%%%%%%%%%%%%%%%%%%%%%%%%%%%%%%%%%%%%%%%%
This section begins by reviewing the general structure of correlation functions of arbitrary spin operators for any CFT${}_3$ \cite{Giombi:2011rz} and proceeds by reviewing the construction of boundary-to-bulk propagators for massless fields in AdS${}_4$, see e.g. \cite{Giombi:2010vg,Didenko:2012tv,Sezgin:2017jgm}. In the last part we generalize the results of \cite{Sezgin:2017jgm} as to compute holographic correlation functions for any contact interaction vertex in the bulk. The formalism makes the helicity structure manifest, as will be explored in the next section. 

%%%%%%%%%%%%%%%%%%%%%%%%%%%%%%%%%%%%%%%%%%%%%%%%%%%%%%%%%%%%%%%%%%%%%%%
\subsection{General Structure of Correlators} 
%%%%%%%%%%%%%%%%%%%%%%%%%%%%%%%%%%%%%%%%%%%%%%%%%%%%%%%%%%%%%%%%%%%%%%%
In the CFT context we are interested in the general structure of conformally-invariant correlation functions of operators that can carry a nontrivial representation of the Lorentz algebra. A very useful observation \cite{Costa:2011mg} is that any correlation function can always be expressed as a sum over a finite number of functions of cross-ratios, each of which is multiplied by a monomial in certain conformally-invariant structures that takes care of the Lorentz spin of the operators. For $d=3$, on one hand, the structure of these conformally invariant building blocks of (spin)-tensors can be greatly simplified \cite{Giombi:2011rz} with the help of the vector-spinor dictionary, which manifests $so(2,1)\sim sl(2,\mathbb{R})$. On the other hand, there are new features --- parity odd structures enter the game early on: there can be parity breaking contact contributions to two-points functions, which we ignore by working with point split correlators; there can be also be parity odd structures in three-point functions, which are important for the bosonization duality to take place \cite{Giombi:2011rz,Maldacena:2011jn}. 

In $3d$ an irreducible tensor has to be totally symmetric, i.e. no mixed-symmetry option, and traceless. Therefore, quasi-primary operators $O^{a_1...a_s}_\Delta$ are required to be such, as well as to have a certain conformal weight $\Delta$. Thanks to $so(1,2)\sim sl(2,\mathbb{R})$ we can replace $x^m$, $m=1,2,3$ with a symmetric bispinor, $\rmx^{\ga\gb}=\rmx^{\gb\ga}$. A traceless rank-$s$ tensor is mapped into a rank-$2s$ symmetric spin-tensor. To save our eyes from seeing too many indices, we prefer to contract them with an auxiliary polarization spinor
$\eta\equiv \eta^\ga$, which leads to the following transmutations
\begin{align}
O^{a_1...a_s}_\Delta(\rmx) \qquad \longrightarrow \qquad O^{\alpha_1...\alpha_{2s}}_\Delta(\rmx^{\gb\gc}) 
\qquad \longrightarrow 
\qquad  
O_\Delta(\rmx,\eta)=O^{\alpha_1...\alpha_{2s}}_\Delta(\rmx)\eta_{\alpha_1}...\eta_{\alpha_{2s}}\ .
\end{align}
It is clear that a correlation function of some operators $O_{\Delta_i}(\rmx_i,\eta_i)$, which are inserted at points 
$\rmx_i$ with all indices contracted with $\eta^i_\ga$, is a conformally-invariant function of $\rmx_i,\eta_i$. For example, the Lorentz transformations are represented by $SL(2,\mathbb{R})$ matrices $A\fdu{\ga}{\gb}$ that acts as
\begin{align}
\rmx^{\gb\gd}&\rightarrow A\fdu{\ga}{\gb}A\fdu{\gc}{\gd}\rmx^{\ga\gc} \ ,
&& \eta^i_\ga \rightarrow A\fdu{\ga}{\gb} \eta^i_\gb\ .
\end{align}
The inversion map $R$ plays an important role in what follows and is defined as 
\begin{align}
R\vec\rmx &=\frac{\vec \rmx}{\rmx^2}\ , & 
R\eta_\alpha &=\frac{\rmx_\alpha{}^\beta \eta_\beta} {\rmx^2}\ . 
\label{prime}
\end{align}
A (over)complete set of conformal invariants that encodes the Lorentz properties of operators is given \cite{Giombi:2011rz} by
\begin{align}
P_{ij}&=  \eta^i_\ga R[\rmx_i-\rmx_j]^{\ga\gb}\eta^j_\gb\ , && RP_{ij}=P_{ij}\ ,
\\
Q^i_{jk}&= \eta^i_\ga \left(R[\rmx_j-\rmx_i]-R[\rmx_k-\rmx_i]\right)^{\ga\gb}\eta^i_\gb\ , 
&& RQ^i_{jk}=Q^i_{jk} \,,
\end{align}
which are parity even, and by a parity-odd structure
\begin{align}
S^i_{jk} =  \frac{\eta^k_\ga (\rmx_{ki})\fud{\ga}{\gb}(\rmx_{ij})^{\gb\gc} \eta^j_\gc}{\rmx_{ij} \rmx_{ik} \rmx_{jk}}\ , 
&& RS^i_{jk}=-S^i_{jk}\ .
\end{align}
Any three-point correlation function 
$\langle O_1(\rmx_1,\eta^1)O_2(\rmx_2,\eta^2) O_3(\rmx_3,\eta^3)\rangle $ can be decomposed 
into an obvious prefactor times a polynomial in $Q,P,S$ structures:
\begin{align}\label{genstr}
\langle O_1(\rmx_1,\eta^1)O_2(\rmx_2,\eta^2) O_3(\rmx_3,\eta^3)\rangle &= 
\frac{1}{\rmx_{12}^{\tau_1+\tau_2-\tau_3} \rmx_{13}^{\tau_1+\tau_3
-\tau_2}\rmx_{23}^{\tau_2+\tau_3-\tau_1}}f(P,Q,S)\,,
\end{align}
where $\tau=\Delta-s$ is the twist. 
Function $f$ has to agree with the value of spin at every point, which imposes simple restrictions. For spins high enough there is a lot of ambiguity hidden in $f$ due to algebraic identities among $P$, $Q$ and $S$, \cite{Giombi:2011rz}. We will deal with three-point functions and it is convenient to introduce the following abbreviations
\begin{align}
Q_1&\equiv Q^1_{32}\ , &Q_2&\equiv Q^2_{13}\ , & Q_3&\equiv Q^3_{21}\ , &
S_1&\equiv S^1_{32}\ , &S_2&\equiv S^2_{13}\ , & S_3&\equiv S^3_{21}\,,
\end{align}
and we also add $P_3\equiv P_{12}$, $P_1\equiv P_{23}$, $P_2\equiv P_{31}$. Altogether, the variables so defined go into each other under the cyclic permutations of $123$. For the algebraic relations among these nine structures, see Appendix \ref{app:notation}. The even structures $P,Q$ appear in the simplest correlation functions that are completely fixed by the conformal symmetry:
\begin{align}
\langle J_{s_1}(\rmx_1,\eta_1) J_{s_2}(\rmx_2,\eta_2) \rangle & 
\sim\frac{1}{\rmx^2_{12}} \delta_{s_1,s_2}(P_{12})^{s_1+s_2}\ ,
\w2
\langle J_{s_1}(\rmx_1,\eta_1) J_0(\rmx_2) J_0(\rmx_3)\rangle &
\sim\frac{1}{\rmx_{12}\rmx_{23}\rmx_{31}} (Q_1)^{s_1}\ ,
\end{align}
where $J_s$ is the spin-$s$ conserved tensor and the weight of scalar 
operator $J_0$ is $\Delta=1$. The conservation of $J_s$, which is AdS/CFT dual to the masslessness in AdS, plays an important role in the paper and can be checked/imposed with the help of
\begin{align}
\mathrm{div} = \frac{\pl^2}{\pl \eta_\ga  \pl \eta_\gb}\frac{\pl}{ \pl \rmx^{\ga\gb} }\ .
\end{align}

\paragraph{Three-point functions and bosonization.}
The main kinematical statement about the general structure of three-point functions of higher spin currents \cite{Giombi:2011rz,Giombi:2011kc,Maldacena:2012sf} is that\footnote{The existence of the odd CFT${}_3$ structure was pointed out in \cite{Maldacena:2011nz}. }
\begin{align}\label{threept}
    \langle J_{s_1} J_{s_2} J_{s_3}\rangle&= a_{s_1,s_2,s_3}\langle J_{s_1} J_{s_2} J_{s_3}\rangle_{\text{F.B.}}+ b_{s_1,s_2,s_3}\langle J_{s_1} J_{s_2} J_{s_3}\rangle_{\text{F.F.}}+ c_{s_1,s_2,s_3}\langle J_{s_1} J_{s_2} J_{s_3}\rangle_{\text{odd}}\,,
\end{align}
where the first two structures can be computed in the theories of free boson and free fermion. The last one does not show up in any of the free theories and is parity violating. The coefficients $a,b,c$, which are related to the OPE coefficients, can be spin-dependent and encode the dynamical information about a given CFT. The free fermion and free boson structures are conserved for every point. However, the parity odd structure is conserved for every leg only for spins that satisfy triangle inequalities, i.e. one can draw a triangle with edges of lengths $s_{1,2,3}$. For spins that cannot form the edges of a triangle, e.g. $s_1>s_2+s_3$, the conservation can be imposed on $J_{s_2}$ and $J_{s_3}$, but not on $J_{s_1}$. This fact has a simple explanation: in interacting theories higher spin currents must not be conserved, but the non-conservation operator, which stays on the r.h.s.\ of $\pl \cdot J$, has to be another quasi-primary operator to the leading order in the large-$N$ expansion. By counting spins and dimensions of the operators that are available in vector models, it is easy to see that the non-conservation is driven by the higher spin currents themselves. Schematically, one finds\footnote{All $a$ indices are to be symmetrized, traces projected out and $a(k)\equiv a_1...a_k$ denotes a group of symmetric indices.}
\begin{align}\label{sbhs}
    \pl^m J_{ma(s-1)}&= \sum_{s_1,s_2,... } C_{s_1,s_2}^{s}[J_{s_1} J_{s_2}]=\sum_{s_1,s_2,... } C_{s_1,s_2}^{s}[\pl_{a}...\pl_a J_{a(s_1)} \pl_a...\pl_a J_{a(s_2)}]\,,
\end{align}
where $C$ are some structure constants containing the dynamical information, while various terms in the brackets with same $J_{s_1}$, $J_{s_2}$ but different arrangement of derivatives are fixed by the representation theory in such a way that $[J_{s_1} J_{s_2}]$ is quasi-primary. It is clear that $s_1> s_2+s_3$, which explains why the parity odd structure can (in fact, has to) be non-conserved for spins outside the triangle inequality.  

An important dynamical statement \cite{Maldacena:2012sf} is that the three-point functions are fixed by the slightly-broken higher spin symmetry manifested via \eqref{sbhs}. In a certain normalization for the building blocks in \eqref{threept} the three-point functions must have the form \eqref{threept} with 
\begin{align}
    a&= \tilde{N}\cos^2 \theta\,, & b&= \tilde{N}\sin^2 \theta\,, & c&= \tilde{N}\cos \theta \sin \theta \,,
\end{align}
where $\tilde{N}$ and $\theta$ are certain phenomenological parameters here. Large-$N$ Chern-Simons matter theories satisfy the assumptions to have slightly-broken higher spin symmetry via \eqref{sbhs}. In particular, phenomenological parameters $\tilde{N}$ and $\theta$ can be related to the microscopical ones, the number of fields $N$ and Chern-Simons level $k$ in concrete theories. $\tilde{N}$ is irrelevant for us as it depends on the overall normalization of correlators. It is important that there is one essential parameter $\theta$ and it enters in a very simple trigonometric way.\footnote{Based on the realization of the slightly-broken higher spin symmetry via a certain strong homotopy algebra \cite{Sharapov:2018ioy,Sharapov:2020quq} it can be proved that all $n$-point correlation functions are made of several fixed structures multiplied by powers of $e^{\pm i\theta}$, i.e. cannot be any complicated dependence on the phenomenological couplings \cite{Gerasimenko:2021sxj}.} There is a similar statement for three-point functions involving one scalar operator. Kinematically there are two independent conformally-invariant structures
\begin{align}\label{threeptA}
    \langle J_{s_1} J_{s_2} J_{0}\rangle&= a_{s_1,s_2,0}\langle J_{s_1} J_{s_2} J_{0}\rangle_{\text{F.X.}}+ c_{s_1,s_2,0}\langle J_{s_1} J_{s_2} J_{0}\rangle_{\text{odd}}\,,
\end{align}
where $J_0$ is either $\Delta=1$ or $\Delta=2$ scalar operator. In the former case, F.X.\ means F.B.\ and in the latter F.X.\ = F.F.\ is meant. The odd structures can be extracted from the critical vector and Gross-Neveu models. Slightly-broken higher spin symmetry implies (again, with a certain normalization of the building blocks)
\begin{align}
    a&= \tilde{N}\cos\theta\,, & c&= \tilde{N}\sin \theta\,.
\end{align}
The structure of three-point functions with two and three scalar operators is fixed by conformal symmetry. For further convenience we collect several generating functions in the free boson and free fermion theories. We will refer to them later when establishing the dictionary between interactions in AdS${}_4$ and correlators on the boundary.  

\paragraph{Generating functions for Free Boson. } There is a simple generating function for the correlators of higher spin currents in the free boson CFT \cite{Giombi:2011rz,Giombi:2010vg,Colombo:2012jx,Didenko:2012tv}:
\begin{align}
   \langle J J J\rangle_{\text{F.B.}}&= \frac{1}{\rmx_{12}\rmx_{23}\rmx_{31}}\exp\left({\tfrac{i}{2} \left(Q_1+Q_2+Q_3\right)}\right) \cos \left(P_1\right) \cos \left(P_2\right) \cos \left(P_3\right)\,.
\end{align}
It is valid for $s=0$ operator $J_0=\phi^2$ as well. Generating functions for all $n$-point correlators can be found in \cite{Didenko:2012tv}.

\paragraph{Generating functions for Free Fermion. } There are similar generating functions for the free fermion CFT \cite{Giombi:2011rz,Giombi:2010vg,Colombo:2012jx,Didenko:2013bj}, which are valid for $s>0$ and those with $\Delta=2$ operator $\tilde{J}_0$ have to be given separately:
\begin{align}
    \langle J J J\rangle_{\text{F.F.}}&= \frac{1}{\rmx_{12}\rmx_{23}\rmx_{31}}\exp\left({\tfrac{1}{2} i \left(Q_1+Q_2+Q_3\right)}\right) \sin \left(P_1\right) \sin \left(P_2\right) \sin \left(P_3\right)\,,\\
    \langle \tilde J_0 J J\rangle_{\text{F.F.}}&=\frac{1}{\rmx_{12}^2\rmx_{31}^2}\exp\left(\tfrac{1}{2} i \left(Q_2+Q_3\right)\right) S_1 \sin \left(P_1\right)\,.
\end{align}
All $n$-point functions can be found in \cite{Didenko:2013bj}. Let us also give several low spin examples that will be relevant for AdS/CFT applications later on. 

\paragraph{Currents' three-point, $\boldsymbol{1-1-1}$. } There are two parity even and one parity odd structure, if we impose cyclic symmetry over the three legs. As a result, there is a definite symmetry under the other permutations. In this case all structures are anti-symmetric and read:\footnote{We omit the obvious $1/\rmx$ prefactors that are present in \eqref{genstr}. }
\begin{align}
    \langle J_1 J_1 J_1 \rangle_{\text{even}} &=a_1 P_1 P_2 P_3 +a_2 Q_1Q_2Q_3\,,\\
    \langle J_1 J_1 J_1 \rangle_{\text{odd}} &= a_3 (P_1 Q_1 S_1+P_2 Q_2 S_2+P_3 Q_3 S_3)\sim S_1 S_2 S_3\,.
\end{align}
All the three structures are conserved, as can easily be checked. The first one, $P_1 P_2 P_3$ comes from the free fermion CFT. The free boson CFT gives the l.h.s.\ of
\begin{align}
    2 P_1^2 Q_1+2 P_2^2 Q_2+2 P_3^2 Q_3+Q_1 Q_2 Q_3 \equiv 3 Q_1 Q_2 Q_3-4 P_1 P_2 P_3\,.
\end{align}
With the help of the identities, see Appendix \ref{app:identities}, it can be converted into the r.h.s. Similarly, there are two equivalent forms of the odd structure.

\paragraph{Stress-tensors' three-point, $\boldsymbol{2-2-2}$. } There are three independent conformally-invariant structures: free fermion, free boson and an odd one that we choose to be
\begin{align}
        \langle J_2 J_2 J_2 \rangle_{\text{odd}} &=\left(P_1 P_2 P_3+Q_1 Q_2 Q_3\right) \left(P_2 P_3 S_1+P_1 P_3 S_2+P_1 P_2 S_3\right)\,.
\end{align}

%%%%%%%%%%%%%%%%%%%%%%%%%%%%%%%%%%%%%%%%%%%%%%%%%%%%%%%%%%%%%%%%%%%%%%%
\subsection{Bulk-to-boundary propagators for (massless) fields} 
\label{subsec:propagators}
%%%%%%%%%%%%%%%%%%%%%%%%%%%%%%%%%%%%%%%%%%%%%%%%%%%%%%%%%%%%%%%%%%%%%%%
We will mostly deal with massless fields of arbitrary spins. They are AdS/CFT dual to higher spin currents. They can be described by symmetric tensors $\Fron_{\mm_1...\mm_s}(x)$ that obey (in our units of cosmological constant) the Klein-Gordon equation
\begin{align}\label{FronsdalEq}
(\square +2(s^2-2s-2))\Fron_{a_1...a_s}&=0\ , && \Fron\fud{c}{ca_3...a_s}=0\ , 
&& \nabla^c\Fron_{ca_2...a_s}=0\,,
\end{align}
together with the tracelesness and transversality constraints. The latter is a result of partial gauge fixing of $\delta \Fron_{a_1...a_s} = \nabla _{(a_1} \xi_{a_2...a_s)}$. The value of the mass-like term is determined by the gauge invariance. To get a massive spin-$s$ field one should replace the fixed mass of the massless field in \eqref{FronsdalEq} with a free parameter $m^2$. We will often use the dictionary between world and fiber tensors established with the help of AdS${}_4$ vierbein $h^a\equiv h^a_\mm \, dx^\mm$. 

Taking advantage of the isomorphism $sl(2,\mathbb{C})_{\mathbb{R}}\sim so(1,3)$, we can replace every vector index $m=0,...,3$ with bi-spinor $x^{\ga\gad}\equiv \sigma_m^{\ga\gad} x^m$. Likewise, a traceless symmetric rank-$s$ tensor $\Fron^{a_1...a_s}$ becomes $\Phi^{\ga(s),\gad(s)}$, where we introduced a shorthand $\ga(s)$ for a group of symmetric indices $\ga_1...\ga_s$. Analogously, vierbein $h^a$ morphs into $h^{\ga\gad}$, see Appendix \ref{app:notation} for more notation.

\paragraph{Complete basis of geometric structures for the boundary-to-bulk problem.} We would like to introduce a complete set of objects with a clear geometric interpretation that allows us to express any structure relevant for the boundary-to-bulk problem in what follows \cite{Didenko:2012vh,Didenko:2012tv}. Suppose there is an operator $O_\Delta^{\ga(2s)}(\rmx)$ on the boundary and a spin-$s$ field $\Phi^{\ga(s),\gad(s)}(x)$, where the coordinates of the bulk point $x^{\ga\gad}$ can be identified, as a two-by-two matrix, with $\rmx^{\ga\gb}+ i\epsilon^{\ga\gb} z$.

The first geometric object is $K_a$, which is an analog of the geodesic distance between the bulk point $(\rmx,z)$ and boundary point $\rmx_a$ (it can be understood as a limit of a function of the geodesic distance $d(x,y)$ whenever one point approaches the boundary):
\begin{align}
K_a&=\frac{z}{(\rmx-\rmx_a)^2+z^2}\,. %\qquad K_0 \equiv K\ ,
\end{align}
It is the same as the bulk-to-boundary propagator for the scalar field that is dual to $\Delta=1$ scalar operator. The bulk/boundary translation invariance in the $\rmx$-direction allows us to set $\rmx_a=0$ in this section for simplicity. The geodesic distance $K$ can be used to define a wave-vector  $\tanV^{\ga\gad}$ (it can be understood as a limit of the tangent vector to the geodesic):
\begin{align}
d K&=K \tanV_{\ga\gad} h^{\ga\gad}\ ,
\label{AppbtobB}
\end{align}
where
\begin{align}\label{bloodyF}
\tanV^{\ga\gad}&=\left(\frac{2z}{\rmx^2+z^2}\rmx^{\ga\gad}-
\frac{\rmx^2-z^2}{\rmx^2+z^2}\,i\epsilon^{\ga\gad}\right)\ .
\end{align}
To relate the tangent spaces at the bulk and boundary points we need the parallel-transport bi-spinors $\Pi^{\ga\gb}$ and 
$\Pib^{\gad\gb}$ given by
\begin{align}\label{Xidef}
\Pi^{\ga\gb}&=K \left(\frac{1}{\sqrt{z}}\,
\rmx^{\ga\gb}+\sqrt{z}\,i\epsilon^{\ga\gb}\right)\ , &
\Pib^{\gad\gb}&=K \left(\frac{1}{\sqrt{z}}\,
\rmx^{\gad\gb}-\sqrt{z}\,i\epsilon^{\gad\gb}\right)=(\Pi^{\ga\gb})^\dag\,.
\end{align}
They should be used to propagate the boundary polarization spinors $\eta^\ga$ into the bulk:
\begin{align}
%\label{Xidef}
\xi^\ga &= \Pi^{\ga\gb}\eta_\gb\,e^{+i\frac{\pi}4} \ , \qquad 
\bar{\xi}^{\gad}=(\xi^\ga)^\dag=\Pib^{\gad\gb}\eta_\gb\,e^{-i\frac{\pi}4} \ .
\end{align}
This is the full set of the data that any geometric quantity, including the bulk-to-boundary propagator, can depend on. The set is closed 
under covariant derivatives and algebraic manipulations, see Appendix \ref{app:identities} for various identities. 

\paragraph{Propagators.}
With the help of the geometric objects just introduced, given the irreducibility of $\Phi^{\ga(s),\gad(s)}$, there is a unique expression for the boundary-to-bulk propagator of a massless field:
\begin{align}\label{propagator}
\Fron_{\ga(s),\gad(s)}&=b_s K \xi_{\ga(s)} \brxi_{\gad(s)} \ ,
\end{align}
where $b_s$ is an a priory arbitrary normalization and $\xi_{\ga(s)}\equiv \xi_{\ga_1}...\xi_{\ga_s}$. We stick to $b_s=1$. The only modification required to get the propagator for a massive spin-$s$ field is to change $K$ to $K^\nu$ for an appropriate relation between $\nu$ and $m^2$. 

In practice, most of the interactions involve a number of derivatives. It is not necessary to know all possible derivatives of the on-shell fields as it turns out that the interactions can always be written in such a form that derivatives form the maximal number of curls.\footnote{This is a general statement valid for any spin and in any dimension, see e.g. \cite{Grigoriev:2020lzu}, where this result is formulated in terms of the minimal model of the jet extension of the BRST complex. } A specific feature of $4d$ is that any curl, being a covariant derivative that is anti-symmetrized with an index of the field, e.g. $F_{\mu\nu}=\nabla_\mm A_\nn-\nabla_\nn A_\mm$, can be decomposed into self-dual and anti-self-dual components. This split can nicely be done in the spinorial language. For example, the Maxwell tensor $F_{\mm\nn}$ splits into $F_{\ga\gb}$ and $F_{\gad\gbd}$ as
\begin{align}
    F_{\ga\gb}&= \tfrac12 \nabla\fdu{(\ga}{\gnd}A_{\gb)\gnd}=  \tfrac12 K\xi_\ga\xi_\gb\,, & F_{\gad\gbd}&= \tfrac12 \nabla\fud{\gnd}{(\gad}A_{\gnd\gbd)}=  \tfrac12 K\brxi_\gad\brxi_\gbd \,,
\end{align}
where $A_\mm = A_{\ga\gad} h^{\ga\gad}_\mm$. According to the identities of Appendix \ref{app:identities}, by taking curls one generates a set of fields $\Phi_{\ga(s+n),\gad(s-n)}$ where the curls decrease/increase the number of $\xi$/$\brxi$
\begin{align}
    &K\xi_{\ga(2s)} && K\xi_{\ga(2s-1)}\brxi_\gad && ... && K\xi_{\ga(s)}\brxi_{\gad(s)} && ... && K\xi_\ga\brxi_{\gad(2s-1)} && K\brxi_{\gad(2s)} \,.
\end{align}
In general, we have 
\begin{align}\label{allmycurls}
    \Phi_{\ga(s+n),\gad(s-n)} &= \nabla\fdu{\ga}{\gdd}...\nabla\fdu{\ga}{\gdd} \Phi_{\ga(s),\gad(s-n) \gdd(n)}\,.
\end{align}
Taking the last curl makes $\Phi$ carry only dotted $\Phi_{\gad(2s)}$ or undotted $\Phi_{\ga(2s)}$ indices. For $s=1$ we get the (anti)-self-dual components of the Maxwell tensor $F_{\mu\nu}$. For $s=2$ one finds the (anti)-self-dual parts of the Weyl tensor. After $s$ curls are taken, which is the maximal number of curls, one can take gradients 
\begin{align}\label{higherder}
    \Phi_{\ga(2s+k),\gad(k)}&= \nabla_{\ga\gad} ...\nabla_{\ga\gad}\Phi_{\ga(2s)}\,, & \Phi_{\ga(k),\gad(2s+k)}&= \nabla_{\ga\gad} ...\nabla_{\ga\gad}\Phi_{\gad(2s)}\,,
\end{align}
where the indices denoted by the same letter are to be symmetrized. Going on-shell, each $\nabla_{\ga\gad}$ produces nothing else but $\tanV_{\ga\gad}$ \eqref{bloodyF}. 

This simple picture is very similar to what one finds in Minkowski space. Indeed, the plane-wave solution for a spin-$s$ field can be written as
\begin{align}
\begin{aligned}
\Phi_{\ga(s),\gad(s)}(p)&=
    \frac{\xi_{\ga_1}...\xi_{\ga_s}\brmu_{\gad_1}...\brmu_{\gad_s}}{(\brxi \brmu)^s} \exp [\pm i x^{\gb\gbd}\xi_\gb \brxi_{\gbd}] \Phi_{+s}(\xi)+\\
    &+\frac{\mu_{\ga_1}...\mu_{\ga_s}\brxi_{\gad_1}...\brxi_{\gad_s}}{(\xi \mu)^s} \exp [\pm i  x^{\gb\gbd}\xi_\gb \brxi_{\gbd}] \Phi_{-s}(\xi)\,.
\end{aligned}
\end{align}
Here, the on-shell momentum $p^m$, $p^2=0$ is factorized into the product of spinors $p^{\ga\gad}=\xi^{\ga}\brxi^{\gad}$, as usual. The reference spinor is $\mu$. Positive and negative helicity eigenstates are represented by $\Phi_{\pm s}$. As is anticipated, $\Phi_{\ga(s),\gad(s)}$ carries both helicities. Let us take the first curl of it
\begin{align}\label{helicityout}
    \pl\fdu{\ga}{\gdd} \Phi_{\ga(s),\gad(s-1)\gdd}&\sim \frac{\xi_{\ga_1}...\xi_{\ga_{s+1}}\brmu_{\gad_1}...\brmu_{\gad_{s-1}}}{(\brxi \brmu)^{s-1}} \exp [\pm i x^{\gb\gbd}\xi_\gb \brxi_{\gbd}] \Phi_{+s}(\xi) \,.
\end{align}
Taking the curl eliminates one of the helicities and the rest of the curls as well as the corresponding infinite family in \eqref{higherder} contain a single helicity. An important conclusion is that by looking at the vertex we can deduce its helicity structure, e.g. taking at least one curl projects onto a definite helicity state, while the presence of $\Phi_{\ga(s),\gad(s)}$ results in the mixed structure. The simplest example is the Yang-Mills vertex that decomposes into $++-$ and $--+$ due to the presence of $A$. Note, see also below, that this helicity separation is less efficient in (A)dS${}_4$.

The spinor-helicity formalism of \cite{Nagaraj:2018nxq,Nagaraj:2019zmk,Nagaraj:2020sji} preserves the bulk Lorentz symmetry explicitly and is perfect for computing amplitudes in AdS${}_4$. However, to establish the holographic dictionary one has to integrate the wave functions of \cite{Nagaraj:2018nxq,Nagaraj:2019zmk,Nagaraj:2020sji} against appropriate wave-packets (intertwining operators) whose one leg is on the boundary. Our formalism was defined with the help of Poincare coordinates and does not maintain bulk Lorentz symmetry on the first sight. Nevertheless, the composite objects $K$, $\xi_\ga$, $\brxi_\gad$, $\tanV_{\ga\gad}$ obey a set of Lorentz covariant relations. Therefore, the formalism incorporates both a close relation to the boundary correlators as in \cite{Maldacena:2011nz} and bulk Lorentz invariance as in \cite{Nagaraj:2018nxq,Nagaraj:2019zmk,Nagaraj:2020sji}. 

One important property that the present formalism cannot achieve, as we will see later, is to separate positive and negative helicities for components in \eqref{allmycurls} with $n\neq \pm s$. This is a consequence of the fact that \eqref{helicityout} fails to be true in (A)dS${}_4$. Indeed, $[\nabla_{\ga\gad},\nabla_{\gb\gbd}]\xi_\nu\sim \xi_\ga \epsilon_{\gb\nu}\epsilon_{\gad\gbd}+\xi_\gb \epsilon_{\ga\nu}\epsilon_{\gad\gbd}$, which implies that one can always go back from $\Phi_{\ga(s+k),\gad (s-k)}$, $0<k<s$ to any of $\Phi_{\ga(s-i),\gad (s+i)}$, $-k\leq i \leq s$. In other words, any of $\Phi_{\ga(s\pm k),\gad (s\mp k)}$ with $0\leq |k|<s$ can be used to recover the rest and, hence, contains both helicities. Note, however, that all of \eqref{higherder}, starting from $\Phi_{\ga(2s)}$ and $\Phi_{\gad(2s)}$ contain a single helicity. On one hand this non-separability does not prevent us from getting all the required interactions to make contact with the $3d$ bosonization duality, see also an explanation in section \ref{sec:Bosonization}, but on the other hand it is possible to separate helicities within the approach of \cite{Nagaraj:2018nxq,Nagaraj:2019zmk,Nagaraj:2020sji} and it would be interesting to see how to improve the present approach to achieve that. For example, this separation is clearly visible in \cite{Krasnov:2021nsq,Sharapov:2022awp}. We should also note that our AdS spinor-helicity expressions get mapped via AdS/CFT into correlators in the usual spin-tensor language of \cite{Giombi:2011rz}, reviewed in section \ref{sec:review}, and, hence, do not carry over any spinor-helicity meaning to the CFT side. Therefore, comparison with the results obtained via the CFT analog of the spinor-helicity language, e.g. \cite{Jain:2021vrv,Jain:2021gwa,Jain:2021whr}, can done at the level of final correlators only.

%%%%%%%%%%%%%%%%%%%%%%%%%%%%%%%%%%%%%%%%%%%%%%%%%%%%%%%%%%%%%%%%%%%%%%%
\subsection{Holographic correlators}
\label{sec:holocoralg}
%%%%%%%%%%%%%%%%%%%%%%%%%%%%%%%%%%%%%%%%%%%%%%%%%%%%%%%%%%%%%%%%%%%%%%%
The goal of this section, which is mostly delegated to Appendix \ref{app:algorithm}, is to review and generalize the results of \cite{Sezgin:2017jgm}, which are themselves based on the old trick \cite{Freedman:1998tz} of using the inversion isometry of (anti)-de Sitter space. The main challenge is to scalarize the integrand. It is obvious that all possible contact interaction vertices 
\begin{align}
    \mathcal{S}_n& =\int V_n\,, & V_n&= f[K_i, \xi_i, \tanV_i]=\prod_i K_i^{\Delta_i}f[(\xi_i \xi_j), (\brxi_i \brxi_j), (\xi_i \tanV_j \brxi_i), (\xi_i \tanV_j \brxi_k)]
\end{align}
can be expressed in terms of various singlet contractions
\be
(\xi_i\xi_j) \equiv \xi_i^\alpha \xi_{j\alpha}\ ,\qquad \qquad  (\xi_i \tanV_j \bar{\xi}_k)
\equiv \xi_{i\ga} \tanV^{\ga\gad}_j \xi_{k\gad}\ ,
\ee
where 
\begin{align*}
K_i &=K(\rmx-\rmx_i,z)\ , &
\tanV^{\ga\gad}_i &=\tanV^{\ga\gad}(\rmx-\rmx_i,z)\ , &
\xi^{\ga}_i &=\xi^{\ga}(\rmx-\rmx_i,z;\eta_i)\ , &
\bar{\xi}^{\gad}_i &=\bar{\xi}^{\gad}(\rmx-\rmx_i,z;\eta_i) \ ,
\end{align*}
refer to the $i$-th boundary point. 
Not all of the structures are independent due to the usual freedom to integrate by parts and also due to Fierz identities. In order to have (an overcomplete) basis we can choose $(\xi_i\xi_j)$, $i<j$ and complex conjugates thereof. As for $(\xi_i \tanV_j \bar{\xi}_k)$, we should have $j\neq i,k$ (otherwise, we can use some of the algebraic identities to reduce it to $(\xi_i\xi_k)$-type, see Appendix \ref{app:identities}). We should also have in mind that $(\xi_i \tanV_j \bar{\xi}_k)^\dag=(\xi_k \tanV_j \bar{\xi}_i)$. 

The three-point integrals, $\mathcal{S}_3$, are doable in principle due to the fact that one can always `scalarize' the integrand by representing all $x^{\ga\gad}$, which are hidden inside $\xi$'s and $\tanV$'s, as derivatives with respect to the boundary points $\rmx_i$. With the help of the $3d$ translation invariance we set $\rmx_1=0$ and then use the inversion map. As a result all the building blocks of the bulk vertices drastically simplify and we can scalarize any integrand (even for $n$-point vertices, see Appendix \ref{app:npoint}). The integral is then trivially computed and the derivatives resulted from the scalarization procedure are brought back. At the final step one recovers the conformally invariant structures $P$, $Q$ and $S$. The algorithm is explained in Appendix \ref{app:algorithm}. Let us present below some simple examples that recover all the basic conformally-invariant structures from simple vertices in the bulk
\begin{subequations}
\begin{eqnarray}
\int \frac{d^{3}\mathrm{x}dz}{z^{4}}K_{i}K_{j}K_{k}\left( \xi _{j}\xi _{k}+%
\bar{\xi}_{j}\bar{\xi}_{k}\right)  &=&\frac{\pi ^{3}}{2}\frac{1}{\mathrm{x}%
_{ij}\mathrm{x}_{jk}\mathrm{x}_{ki}}P_{jk}\,, \\
\int \frac{d^{3}\mathrm{x}dz}{z^{4}}K_{i}K_{j}K_{k}\left( \xi _{j}\xi _{k}-%
\bar{\xi}_{j}\bar{\xi}_{k}\right)  &=&i\pi ^{2}\frac{1}{\mathrm{x}_{ij}%
\mathrm{x}_{jk}\mathrm{x}_{ki}}S_{kj}^{i}\,, \\
\int \frac{d^{3}\mathrm{x}dz}{z^{4}}K_{i}^{2}K_{j}K_{k}\left( \xi _{j}\xi
_{k}-\bar{\xi}_{j}\bar{\xi}_{k}\right)  &=&\frac{i\pi ^{3}}{4}\frac{1}{%
\mathrm{x}_{ij}^{2}\mathrm{x}_{ki}^{2}}S_{kj}^{i}\,, \\
\int \frac{d^{3}\mathrm{x}dz}{z^{4}}K_{i}K_{j}K_{k}\left( \xi _{j}\tanV_{k}\bar{%
\xi}_{j}\right)  &=&\frac{\pi ^{3}}{8}\frac{1}{\mathrm{x}_{ij}\mathrm{x}_{jk}%
\mathrm{x}_{ki}}Q_{ik}^{j}\,, \\
\int \frac{d^{3}\mathrm{x}dz}{z^{4}}K_{i}K_{j}K_{k}\left( \xi _{i}\tanV_{j}\bar{%
\xi}_{k}\right)  &=&0\,,  \\
\int \frac{d^{3}\mathrm{x}dz}{z^{4}}K_{i}^{2}K_{j}K_{k}^{2}\left( \xi
_{i}\tanV_{j}\bar{\xi}_{k}\right)  &=&\frac{i\pi ^{2}}{8}\frac{1}{\mathrm{x}_{ij}%
\mathrm{x}_{jk}\mathrm{x}_{ki}^{3}}S_{ik}^{j}\,,
\end{eqnarray}
\end{subequations}
where we assume that whenever three indices are present $i,j,k$, they are all different, $i\neq j\neq k \neq i$. From the dictionary we constructed in Appendix \ref{app:algorithm} it looks like the following variables, which are eigen states of the parity operator, are more convenient:
\begin{align}
    \xi_{ij}^\pm&= \tfrac12[ (\xi_i \xi_j) \pm (\brxi_i \brxi_j)] \,.
\end{align}
There is nothing to scalarize for $\xi^+_{ij}$ and it gets mapped to $P_{ij}$ for any number of legs. More examples will be given in due time.

%%%%%%%%%%%%%%%%%%%%%%%%%%%%%%%%%%%%%%%%%%%%%%%%%%%%%%%%%%%%%%%%%%%%%%%
\section{General structure of amplitudes}
\label{sec:amplitudes}
%%%%%%%%%%%%%%%%%%%%%%%%%%%%%%%%%%%%%%%%%%%%%%%%%%%%%%%%%%%%%%%%%%%%%%%
Every CFT${}_3$ structure can be encoded in AdS${}_4$ as a certain vertex. As far as conserved currents and tensors are the objects of interest, it will be clear that AdS${}_4$ vertices provide a very economical way to encode CFT${}_3$ correlation functions, at least, at the $3$-point level\footnote{The AdS-duals of vector models do not have a large gap in the spectrum and, for this reason, have to be very non-local \cite{Bekaert:2015tva,Maldacena:2015iua,Sleight:2017pcz,Ponomarev:2017qab}, more non-local than the field theory approach allows for. } and, at least, for conserved tensors. Technically what happens is that correlation functions of conserved currents/tensors correspond to nontrivial polynomials in $P$, $Q$, $S$, which is due to the fact that the conservation condition does not have a simple solution in position space.\footnote{Various three-point amplitudes in the low spin theories were computed in a version of the spinor-helicity formalism in momentum space, see e.g. \cite{Maldacena:2011nz,Raju:2012zs}. However, to the best of our knowledge neither the computation in position space was performed, nor the relation to the bosonization duality uncovered, except for \cite{Giombi:2011kc,Giombi:2012ms}. } On the contrary, any gauge invariant vertex of massless fields in AdS${}_4$ automatically leads to a correlation function that is conserved unless the contribution of boundary terms is important. Indeed, in checking gauge invariance one has to integrate by parts and there can be nonvanishing contributions from the boundary terms, which is what happens for some vertices/correlators.

\paragraph{Vertices vs. Correlators in the helicity base. } A complete classification of AdS${}_4$ cubic vertices of massless fields was obtained by Metsaev in \cite{Metsaev:2018xip} within the light-front approach to dynamics. Due to the fact that light-cone gauge and spinor-helicity language are tightly connected, the classification of three-point amplitudes within the latter \cite{Nagaraj:2018nxq,Nagaraj:2019zmk,Nagaraj:2020sji} is perfectly consistent with the former. A very special feature of four dimensions is that massless fields with spin have two degrees of freedom that can be represented by two `scalars' $\Phi_{\pm s}$ that are helicity eigen states and are each other's complex conjugates. As a result, every vertex in four dimensions can be decomposed into pieces, $\VVV^{\lambda_1,...,\lambda_n}$, each having a definite helicity structure, e.g. for cubic vertices we can have structures ranging from $+++$ to $---$.  

One of the consequences of \cite{Metsaev:2018xip} is that there is one-to-one correspondence between cubic vertices in flat $M_4$ and AdS${}_4$ spaces. For this reason let us recall the classification of cubic vertices of massless fields in Minkowski space \cite{Bengtsson:1986kh}, see also \cite{Metsaev:1991mt,Metsaev:1991nb,Benincasa:2011pg}. Given any triplet of helicities $h_1$, $h_2$, $h_3$ one can construct the following amplitudes\footnote{To ease the notation we are using the same variables $\xi_i$, $\brxi_i$ for the spinors representing the on-shell momentum at leg $i$, $p^{\ga\gad}_i=\xi^\ga_i \brxi^{\gad}_i$ in flat space. The brackets mean the scalar product with the help of $\epsilon_{\ga\gb}$, $\epsilon_{\gad\gbd}$.}
\begin{align}\label{flatamplitudes}
  h_1+h_2+h_3<0&: &&  \VVV^{h_1,h_2,h_3}\sim\langle \xi_1 \xi_2\rangle^{h_1+h_2-h_3}\langle\xi_2 \xi_3\rangle^{h_2+h_3-h_1}\langle\xi_3 \xi_1 \rangle^{h_3+h_1-h_2}\,,\\
  h_1+h_2+h_3>0&: && \VVV^{h_1,h_2,h_3}\sim[\brxi_1 \brxi_2]^{-(h_1+h_2-h_3)}[\brxi_2 \brxi_3]^{-(h_2+h_3-h_1)}[\brxi_3 \brxi_1]^{-(h_3+h_1-h_2)}\,.
\end{align}
The scalar cubic self-interaction is an exception to this rule and is a unique vertex for $h_{1,2,3}=0$. Therefore, cubic vertices/amplitudes can be identified unambiguously by their helicity structure. The constraint for the sum of helicities to be positive/negative is a consequence of locality. In the light-cone gauge the Hamiltonian has $|h_1+h_2+h_3|$ powers of the transverse momenta:
\begin{align}
  \VVV^{h_1,h_2,h_3}\sim (p_\perp)^{|h_1+h_2+h_3|} \Phi_{h_1}\Phi_{h_2}\Phi_{h_3}  \ .
\end{align}
The boost generators $J^{i-}$ should have one power less and it has to be non-negative, which implies $|h_1+h_2+h_3|>0$, except for $h_i=0$ where $J^{i-}=0$.

As \cite{Metsaev:2018xip} shows, all vertices in Minkowski space $M_4$ can be uplifted to AdS${}_4$. The curvature effect is in that the leading terms of the vertices get corrected by a tail of lower derivative terms.\footnote{It is important to stress that the classification reviewed above is complete. However, there are (can be) other approaches where an unfortunate choice of covariant field variables may prevent one from realizing all of the amplitudes by local vertices in spacetime, see e.g. discussion at the end of \cite{Krasnov:2021nsq} for the summary and \cite{Conde:2016izb} for the subtleties in flat space. Of course, this has to be true for $n$-point vertices as well. The reason to stress that is due to a confusion in some old literature that claimed that certain vertices do not exist in flat space and can only be constructed in anti-de Sitter, which already contradicts \cite{Bengtsson:1986kh,Metsaev:1991mt,Metsaev:1991nb} and is an outcome of some specific choice of covariant field variables. } Yet another issue is that CPT, which is almost automatic in covariant approaches, prescribes certain specific linear combinations of vertices, while in the helicity base they all look independent. For example the Yang-Mills or Einstein-Hilbert cubic vertices consist of a sum of $\VVV^{+s,+s,-s}$ and $\VVV^{-s,-s,+s}$ for $s=1,2$. It is this sum that can be written in terms of $A_\mu$ or $g_{\mu\nu}$, while we cannot isolate each of the two constituents as local functions of $A_\mu$ and $g_{\mu\nu}$. 

A more fine-grained decomposition of interactions that allows one to isolate pieces with definite helicity structure can be achieved by representing complete theories, e.g. Yang-Mills and gravity, as perturbations of self-dual truncations thereof. Indeed, self-dual theories, e.g. Yang-Mills and gravity in flat space, have only vertex of type $\VVV^{+s,+s,-s}$ (or the opposite), $s=1,2$, \cite{Siegel:1992wd,Chalmers:1996rq}. This vertex, being `half' of the Yang-Mills or Einstein-Hilbert vertex, can be written in a manifestly Lorentz-invariant way, but with the help of other field variables (as compared to $A_\mu$ and $g_{\mu\nu}$). 

To summarize the subtle difference between various approaches, depending on the choice of covariant field variables, we have several options: (a) $\VVV^{\lambda_1,\lambda_2,\lambda_3}$ can be written down in a covariant form; (b) certain vertices $\VVV^{\lambda_1,\lambda_2,\lambda_3}$ may turn out to be inaccessible, i.e. cannot be covariantized with this choice; (c) certain specific linear combinations of $\VVV^{\lambda_1,\lambda_2,\lambda_3}$ with the same spins but different helicities can be accessible. We will see that the spinorial variables introduced in section \ref{subsec:propagators} give an access to some vertices (a), miss some particular ones (b), but allow one to find all the interactions relevant for the bosonization duality via (c). As it is well-known, see e.g. \cite{Maldacena:2011jn, Maldacena:2012sf,Giombi:2011ya}, there is an important difference for $\VVV^{\lambda_1,\lambda_2,\lambda_3}$ such that the spins $s_i=|\lambda_i|$ can form a triangle, i.e. $s_i<s_j+s_k$ for all $ijk$, and those triplets where $s_i>s_j+s_k$ for some $ijk$. It is for the latter condition (we call `outside triangle') that the tensor operator dual to the highest spin field in the vertex may not be conserved.  

\paragraph{Three-point amplitudes in $\boldsymbol{\rm{AdS}_4}$. } It is obvious that any on-shell vertex can be expressed as a scalar function of $\xi$'s and $\tanV$'s:
\begin{align}
    V_3&= K_1K_2 K_3f[\xi_i, \tanV_i]=K_1K_2 K_3f[(\xi_i \xi_j), (\brxi_i \brxi_j), (\xi_i \tanV_j \brxi_i), (\xi_i \tanV_j \brxi_k)]\,,
\end{align}
where $K_i$ to the first power indicate that we are dealing with massless fields. We usually drop these factors below. Existence of field-redefinitions and total derivatives imply that one and the same vertex can be written in (infinitely) many ways.\footnote{Given some vertex, there are infinitely many higher and higher derivative clones, see e.g. \cite{Boulanger:2015ova,Skvortsov:2015lja}. This problem is not present on the CFT side. } Total derivatives are given by a multiple of $\sum_i \tanV^i_{\ga\gad}$. There are also Fierz identities that may allow one to rewrite the same vertex in several equivalent ways. Therefore, we would like to find at least one representative of a vertex that corresponds to a given helicity configuration, the simplest one obviously. 

\underline{\textit{Abelian amplitudes inside triangle.}} Let us consider first the case of three spins $s_1$, $s_2$, $s_3$ that can form a triangle, i.e. $s_i+s_j-s_k\geq0$ for all triplets $i\neq j\neq k \neq i$. In this case we do not have to employ $\tanV$'s from \eqref{bloodyF}, which represent derivatives. The actual derivatives turn out to be hidden in the number of $\xi$'s as compared to $\brxi$.\footnote{Note that $\nabla_{\ga\gad} \xi_\gb = \tanV_{\gb\gad}\xi_\ga$, while Appendix \ref{app:identities} tells us that $\tanV\xi=\brxi$ when indices are contracted. Whenever the derivatives form curls rather than symmetrized gradients, we do not see any $\tanV$.} For the abelian vertices we find 
\besubeqs
\begin{align}
  \VVV^{+s_1,+s_2,+s_3}&: &&  (\xi_1 \xi_2)^{s_1+s_2-s_3}(\xi_2 \xi_3)^{s_2+s_3-s_1}(\xi_3 \xi_1)^{s_3+s_1-s_2}\,,\\
  \VVV^{-s_1,-s_2,-s_3}&: &&(\brxi_1 \brxi_2)^{s_1+s_2-s_3}(\brxi_2 \brxi_3)^{s_2+s_3-s_1}(\brxi_3 \brxi_1)^{s_3+s_1-s_2}\,,
\end{align}
\esubeqs
which in terms of on-shell fields \eqref{allmycurls}, \eqref{higherder} corresponds to\footnote{A proposal for parity-odd structures inside the triangle made in \cite{Maldacena:2011jn} is equivalent to these amplitudes. }
\besubeqs\label{abelian}
\begin{align}
    L_3&= \Phi\fud{\ga(s_1+s_3-s_2)}{\gb(s_1+s_2-s_3)}\Phi\fud{\gb(s_1+s_2-s_3)}{\gc(s_2+s_3-s_1)}\Phi\fud{\gc(s_2+s_3-s_1)}{\ga(s_1+s_3-s_2)}\,, \\
    L_3&= \Phi\fud{\gad(s_1+s_3-s_2)}{\gbd(s_1+s_2-s_3)}\Phi\fud{\gbd(s_1+s_2-s_3)}{\gdd(s_2+s_3-s_1)}\Phi\fud{\gdd(s_2+s_3-s_1)}{\gad(s_1+s_3-s_2)} \,.
\end{align}
\esubeqs
Let us note that this class is the only one where we have a definite helicity on every leg in our representation of amplitudes. Abelian vertices outside triangle, which are still $+++$ or $---$, cannot be made pure and contain an admixture of non-abelian vertices $++-$ or $--+$ in our approach and for that reason cease to be abelian. 

\underline{\textit{Non-abelian amplitudes inside triangle.}} These set of vertices involves the usual Yang-Mills and Einstein-Hilbert vertices. For any triplets $s_{1,2,3}$ the corresponding amplitude can be extracted from \cite{Vasiliev:1986bq}\footnote{The exact expression contains certain corrections induced by gauge invariance:
{\scriptsize\begin{align*}
    \sum_l\frac{l! \Gamma \left(-s_1+s_2+s_3\right) \Gamma \left(-l+2 s_2-1\right) \Gamma \left(-l+2 s_3-1\right)}{\Gamma \left(2 s_2-1\right) \Gamma \left(2 s_3-1\right) \Gamma \left(-l-s_1+s_2+s_3\right)}(\xi_1 \xi_2)^{s_1+s_2-s_3}(\xi_2 \xi_3)^{s_2+s_3-s_1-1-l}(\xi_3 \xi_1)^{s_3+s_1-s_2}\xiscb{2}{3}^{1+l} \,.
\end{align*}
It can also be integrated by parts to eliminate $\xiscb{2}{3}$ in the $l=0$ term, so that the leading term looks exactly like in flat space.}
}
\begin{align}
    \VV^{s_1,s_2,s_3}&: &&  (\xi_1 \xi_2)^{s_1+s_2-s_3}(\xi_2 \xi_3)^{s_2+s_3-s_1-1}(\xi_3 \xi_1)^{s_3+s_1-s_2}\xiscb{2}{3}+...\,,
\end{align}
where we indicated the leading term. Our notation $\VV$ signals the failure to project onto a single vertex with definite helicities on external lines. We use $\VV$ instead of $\VVV$ to indicate that the vertex is not pure and contains mixed helicity states. The complex conjugate of this amplitude yields, in fact, the same correlation function. The vertex above is a linear combination of all elementary vertices $\VVV^{\lambda_1,\lambda_2,\lambda_3}$ without $+++$ and $---$, e.g.\ it gives $\VVV^{+2,+2,-2}+\VVV^{-2,-2,+2}$ for $s_{1,2,3}=2$, see also below. 

\underline{\textit{(Non)-Abelian amplitudes outside triangle. I.}} There is a clear problem with any of the AdS${}_4$-formulas above once the spins cannot form a triangle. Surprisingly, the expressions in the light-cone and spinor-helicity languages  display no problem both in flat and (A)dS${}_4$, see e.g.\ \eqref{flatamplitudes}. The covariant approaches reveal certain subtleties in flat space as well, see e.g.\ \cite{Conde:2016izb}. The last vertex that still makes sense even though the triangle degenerates into an interval is 
\begin{align}
    V_3&= \Phi^{\ga(s_1+s_3-s_2)\gb(s_1+s_2-s_3)}\Phi_{\gb(s_1+s_2-s_3)}\Phi_{\ga(s_1+s_3-s_2)} \,,
\end{align}
where $s_2+s_3-s_1=0$. Beyond that they Weyl tensors of spin-$s_{2,3}$ fields do not have sufficiently many indices. Therefore, we need to involve derivatives. However, naive expressions like 
\begin{align}
    V_3&= \Phi^{\ga(s_1+s_3-s_2)\gb(s_1+s_2-s_3)}\Phi_{\gb(s_1+s_2-s_3+k),\gdd(k)}\Phi\fdu{\ga(s_1+s_3-s_2+k),}{\gdd(k)} 
\end{align}
will vanish on-shell upon integration by parts. Here, $\Phi_{\ga(2s+k),\gad(k)}$ are just $\nabla_{\ga\gad}...\nabla_{\ga\gad}\Phi_{\ga(2s)}$. The simplest way to increase the spin of one leg without affecting the others is to add $(\xi_i \tanV_j \brxi_i)$-structures:
\begin{align}
   \VV_3&= (\xi_1 \xi_2)^{2s_2}(\xi_3 \xi_1)^{2s_3} (\xi_1 \tanV_i\xi_1)^{s_1-s_2-s_3} +... \,,
\end{align}
where $i=2,3$, the two forms being equivalent. Therefore, the formally negative powers of $(\xi_2\xi_3)$ get resolved into $(\xi_1 \tanV_i\xi_1)$. Obviously, this is a general rule and can be applied to all the vertices outside triangle, i.e. formally we can always think that the AdS${}_4$-vertices have the same form on-shell as the corresponding amplitudes in the spinor-helicity language, but need to replace negative powers of $(\xi_k \xi_j)$ with positive powers of $(\xi_i \tanV_{j,k}\brxi_i)$ structure. 

More rigorously, complete expressions for this class of vertices can be obtained by taking currents from \cite{Gelfond:2006be} and contracting them with $\Phi_{\ga(s-k),\gad(s+k)}$ for appropriate $k$. Without going into details, given helicity eigen states $\Phi_{\ga(2s_1)}$ and $\Phi_{\ga(2s_2)}$ (or the complex conjugates thereof), one can construct an infinite family of conserved currents
\begin{align}
    J_{\ga(2s_1+2s_2+k),\gad(k)}&=\Phi_{\ga(2s_1)} \underbrace{\nabla_{\ga\gad}...\nabla_{\ga\gad}}_{k}\Phi_{\ga(2s_2)}+...\,.
\end{align}
These currents and the complex conjugates thereof allow us to write the usual Noether current interaction
\begin{align}
    \VV^{+s_1,+s_2,+s_3}&=\Phi^{\ga(2s_1+2s_2+k),\gad(k)} J_{\ga(2s_1+2s_2+k),\gad(k)}\,, \\  \VV^{-s_1,-s_2,-s_3}&=\Phi^{\ga(k),\gad(2s_1+2s_2+k)} J_{\ga(k),\gad(2s_1+2s_2+k)}\,,
\end{align}
where $s_3=s_1+s_2+k$. These vertices have $s_1+s_2+ s_3$ derivatives.  $J$ is built either from $++$ or $--$ helicities and $s_3>s_1+s_2$. The vertices cannot be pure $+++$ or $---$ since they lead to nontrivial contributions to Ward identities (they are not conserved with respect to the highest spin).  

\underline{\textit{Non-abelian amplitudes outside triangle. II.}} Another family of conserved currents from \cite{Gelfond:2006be} can be used to construct the rest of the vertices. Given helicity eigen states $\Phi_{\ga(2s_1)}$ and $\Phi_{\gad(2s_2)}$, i.e. with opposite helicities, one can build conserved tensors 
\begin{align}
    J_{\ga(2s_1+k),\gad(2s_2+k)}&=\Phi_{\ga(2s_1)} \underbrace{\nabla_{\ga\gad}...\nabla_{\ga\gad}}_{k}\Phi_{\gad(2s_2)}+...\,.
\end{align}
The corresponding vertex
\begin{align}
    \VV^{\pm s_1,\mp s_2,\pm s_3}&=\Phi^{\ga(2s_1+k),\gad(2s_2+k)} J_{\ga(2s_1+k),\gad(2s_2+k)}\,, 
\end{align}
is not pure and contains various $++-$ and $--+$ combinations. The vertex has up to $s_1+s_2+k+|s_1-s_2|=s_3+|s_1-s_2|$ derivatives. The amplitude is
\begin{align}
  (\xi_1 \xi_2)^{2s_1}(\brxi_3 \brxi_1)^{2s_2} (\xi_1 \tanV_i\brxi_1)^{s_3-s_1-s_2} +... \,.
\end{align}

\textit{Ward identities and non-conservation.} 
A well-established fact is that abelian amplitudes (those resulting from higher derivative vertices) lead to trivial Ward identities. Indeed, since checking conservation of a holographic correlation function amounts to a gauge variation of the corresponding vertex in the bulk, abelian vertices are manifestly gauge invariant and, hence, are conserved with respect to all legs. Non-abelian amplitudes (those originating from the vertices that are not manifestly gauge invariant, e.g. Yang-Mills or Einstein-Hilbert interactions) lead to nontrivial Ward identities. The non-zero contribution can result from integration by parts in the process of checking gauge invariance and is given by a boundary term. However, Ward identities are easier to analyze in momentum space rather than in position space. In this regard it would be beneficial to establish the dictionary between the position space approach of \cite{Giombi:2011rz} and the momentum space results of \cite{Jain:2020puw,Jain:2021vrv,Jain:2021gwa,Jain:2021whr} and \cite{Metsaev:2018xip,Skvortsov:2018uru}.

%%%%%%%%%%%%%%%%%%%%%%%%%%%%%%%%%%%%%%%%%%%%%%%%%%%%%%%%%%%%%%%%%%%%%%%
\section{Correlation functions}
\label{sec:correlators}
%%%%%%%%%%%%%%%%%%%%%%%%%%%%%%%%%%%%%%%%%%%%%%%%%%%%%%%%%%%%%%%%%%%%%%%
To get a feeling of what bulk vertices get mapped to on the CFT side we begin with a number of simple examples, which we tabulate for everyone's convenience. Even though our formalism covers massive fields with spin, we will consider below massless fields only. The scalar field has either $\Delta=1$ or $\Delta=2$. It is dual to a scalar operator we denote $J_0$ for $\Delta=1$ and $\tilde{J}_0$ for $\Delta=2$. Therefore, the bulk integrals are always of the form
\begin{align}\label{genericV}
    \mathcal{S}_3& =\int \vol V_3\,, & V_3&= K_1 K_2 K_3 f[(\xi_i \xi_j), (\brxi_i \brxi_j), (\xi_i \tanV_j \brxi_i), (\xi_i \tanV_j \brxi_k)]\,,
\end{align}
and it is the structure of $V_3$ that we write down. The result of the integral is a correlation function of the form 
\begin{align}
    \langle J J J\rangle&= \frac{1}{\rmx_{12}\rmx_{23}\rmx_{31}} f(P,Q,S)
    \,, & \langle \tilde{J}_0 J J\rangle&= \frac{1}{\rmx_{12}^2\rmx_{31}^2} f(P,Q,S) \,,
\end{align}
and it is $f$ that we write down. In the $\tilde{J}_0$-case we have $(K_1)^2$ in \eqref{genericV}.

\paragraph{$\boldsymbol{s-0-0}$. } Let us begin with the simplest series of vertices where a scalar field couples to a massless spin-$s$ field via a conserved tensor. After integration by parts and taking advantage of the transverse-traceless gauge we end up with
\begin{align}
    \int \Phi^{a(s)} (\nabla_a... \nabla_a\Phi) \Phi\sim  \int \Phi^{\ga(s),\gad(s)} (\nabla_{\ga\gad}... \nabla_{\ga\gad}\Phi) \Phi \,,
\end{align}
It is clear that the building block is $(\xi_1 F_2 \brxi_1)$, which results in, cf. \cite{Costa:2014kfa}, 
\begin{align}
   \tilde\VVV^{\pm s,0,0}&=\VVV^{+s,0,0}+\VVV^{-s,0,0}: && (\xi_1 F_2 \brxi_1)^n: && \frac{2^{-n-1} \pi ^3\Gamma \left(n+\frac{1}{2}\right)}{\sqrt{\pi } n^2 \Gamma (n)} (Q_1)^n \,.
\end{align}
Using the canonical form of the interaction, as above, it is impossible to separate positive and negative helicities. However, this is exactly the combination that is stable under the correlation functions' surgery \cite{Skvortsov:2018uru} in the helicity basis that manifests the $3d$ bosonization duality. Therefore, we do not attempt to decompose the vertex further, but it may be interesting to see if it is possible to achieve that with the help of a certain seemingly trivial representative, i.e. as a total derivative.

%%%%%%%%%%%%%%%%%%%%%%%%%%%%%%%%%%%%%%%%%%%%%%%%%%%%%%%%%%%%%%%%%%%%%%%
\subsection{Yang-Mills Theory with higher derivative corrections}
\label{subsec:YM}
%%%%%%%%%%%%%%%%%%%%%%%%%%%%%%%%%%%%%%%%%%%%%%%%%%%%%%%%%%%%%%%%%%%%%%%
Our next example offers the simplest possibility to test the spinor-helicity approach and observe parity-odd structures via holographic correlators. The action we take reads
\begin{align}\label{YMaction}
    S_{\text{YM}}&=\tfrac12 \Tr \int \vol \left[ F_{\ga\ga}F^{\ga\ga}+F_{\gad\gad}F^{\gad\gad} \right]\,,
\end{align}
where we define the (anti)-self-dual components of the field strength $F=dA-AA$ as\footnote{For the ease of notation $A$ is assumed to take values in some matrix algebra, so that $AA\equiv dx^\mm\wedge dx^\nn\, A_\mm{}\fud{i}{j} A_\nn{}\fud{j}{k} = \tfrac 12  dx^\mm\wedge dx^\nn\, [A_\mm,A_\nn]\fud{i}{k}.$ It is the trace over these implicit matrix indices that is in front of the action as $\Tr$.}
\begin{align}\label{atoPsi}
    dA-AA&=  E^{\ga\ga} F_{\ga\ga}+  E^{\gad\gad}F_{\gad\gad}\,.
\end{align}
Here $e^{\ga\gad}$ is the (possibly, dynamical) vierbein and we define a basis of (anti)-self-dual two-forms $E^{\ga\ga}\equiv e\fud{\ga}{\gdd}\wedge e^{\ga \gdd}$, $E^{\gad\gad}\equiv e\fdu{\gc}{\gad}\wedge e^{\gc\gad}$. The volume form is $E_{\ga\ga}\wedge E^{\ga\ga}$ and we will mostly use its background AdS${}_4$ value. The AdS${}_4$-vierbein is $h^{\ga\gad}$, see Appendix \ref{app:notation} for more on our notation. 

On top of \eqref{YMaction} we would like to consider two possible higher derivative terms. Their sum is parity even, but a disbalance in the two makes a parity-violating contribution to the CFT correlators. They read
\begin{align}
    &\Tr \int \vol F\fdu{\ga}{\gb}F\fdu{\gb}{\gc}F\fdu{\gc}{\ga}\,, &&  \Tr \int \vol F\fdu{\gad}{\gbd}F\fdu{\gbd}{\gdd}F\fdu{\gdd}{\gad}\,.
\end{align}
Let us recall the structure of the boundary-to-bulk propagators from section \ref{subsec:propagators}. We find
\begin{align}
    A_{\ga\gad}&= K\xi_\ga \brxi_\gad\,, & A&= K h^{\ga\gad}\xi_\ga \brxi_\gad\,, &  F_{\ga\ga}&= \tfrac12K \xi_\ga \xi_\ga\,, & F_{\gad\gad}&= \tfrac12 K \brxi_\gad \brxi_\gad\,,
\end{align}
where the expressions for $F$'s are obtained from \eqref{atoPsi} restricted to free fields. The two terms in the action \eqref{YMaction} are equal modulo a topological term, which is the $\theta$-term
\begin{align}\label{ThetaTerm}
    S_{\text{top}}&=\frac12 \Tr \int \vol \left[ F_{\ga\ga}F^{\ga\ga}-F_{\gad\gad}F^{\gad\gad} \right]\,.
\end{align}
Nevertheless, one should be careful about total derivative since they can affect the answer. Let us work out the cubic amplitude based on the first of them
\begin{align}\label{SDYMa}
    \tfrac12\Tr \int \vol F_{\ga\ga}F^{\ga\ga}\Big|_{\text{cubic}}&= -\tfrac14  \VV^{\mp1,\pm1,\pm1}\,,
\end{align}
where the partial amplitude is defined as\footnote{The trace on the r.h.s. is somewhat superficial. It just implies the color factor and cyclic permutations. }
\begin{align}
    \VV^{\mp1,\pm1,\pm1}=\Tr\int \vol K_1 K_2 K_3 (\xi_1 \xi_2) (\brxi_2\brxi_3)(\xi_3 \xi_1)\,.
\end{align}
This partial amplitude is anti-symmetric in $(23)$ and does not have any other symmetries. 
The second term leads simply to the complex conjugate of the above 
\begin{align}
    \tfrac12\Tr \int \vol F_{\gad\gad}F^{\gad\gad}\Big|_{\text{cubic}}&= -\tfrac14\VV^{\pm1,\mp1,\mp1}\,,
\end{align}
where the amplitude is
\begin{align}
    \VV^{\pm1,\mp1,\mp1}&= \Tr \int \vol K_1 K_2 K_3 (\brxi_1 \brxi_2) (\xi_2\xi_3)(\brxi_3 \brxi_1)\,.
\end{align}
It turns out that $\VV^{\pm1,\mp1,\mp1}=\VV^{\mp1,\pm1,\pm1}$. These two amplitudes correspond to the sum of $\VVV^{-1,+1,+1}$ and $\VVV^{+1,-1,-1}$, which we cannot separate from each other. The higher derivative terms lead to
\begin{align}
    \Tr \int \vol F\fdu{\ga}{\gb}F\fdu{\gb}{\gc}F\fdu{\gc}{\ga}&=\tfrac18 \VVV^{+1,+1,+1}\,, &
    \Tr \int \vol F\fdu{\gad}{\gbd}F\fdu{\gbd}{\gdd}F\fdu{\gdd}{\gad}&=\tfrac18  \VVV^{-1,-1,-1}\,,
\end{align}
where 
\begin{align}
    \VVV^{+1,+1,+1} &=\Tr \int \vol  K_1 K_2 K_3 (\xi_1 \xi_2) (\xi_2\xi_3)(\xi_3 \xi_1)\,,\\
    \VVV^{-1,-1,-1}&=\Tr \int \vol K_1 K_2 K_3 (\brxi_1 \brxi_2) (\brxi_2\brxi_3)(\brxi_3 \brxi_1)\,.
\end{align}
In this case, the amplitudes can be separated. Now, we are ready to compute the holographic correlation functions. The basis to expand is chosen as 
\besubeqs
\begin{align}
    \langle J_1 J_1 J_1 \rangle_{\text{F.B.}} &= \tfrac{1}{8} \left(2 P_1^2 Q_1+2 P_2^2 Q_2+2 P_3^2 Q_3+Q_1 Q_2 Q_3\right)\,, \\
    \langle J_1 J_1 J_1 \rangle_{\text{F.F.}} &=-P_1 P_2 P_3\,,\\
    \langle J_1 J_1 J_1 \rangle_{\text{odd}} &= S_1 S_2 S_3\,.
\end{align}
\esubeqs
The final result can be written as $3\times 3$ matrix:\footnote{Our notation is that $\VV^{(\pm1,\mp1,\mp1)}$ is the sum of $\VV^{\pm1,\mp1,\mp1}$ over all ($3$) cyclic permutations.}
\begin{align}
    \begin{pmatrix}
    \VV^{(\pm1,\mp1,\mp1)}\\
    \VVV^{+1,+1,+1}\\
    \VVV^{-1,-1,-1}
    \end{pmatrix}&= \frac{\pi ^2}{32}
\left(
\begin{array}{ccc}
 -\pi  & -\pi  & 0 \\
 \pi  & -\pi  & -2 i \\
 \pi  & -\pi  & 2 i \\
\end{array}
\right)\begin{pmatrix}
    \langle J_1 J_1 J_1 \rangle_{\text{F.B.}}\\
    \langle J_1 J_1 J_1 \rangle_{\text{F.F.}}\\
    \langle J_1 J_1 J_1 \rangle_{\text{odd}}
    \end{pmatrix}\,.
\end{align}
The inverse transformation is\footnote{Given that both in the bulk and on the boundary the normalization of the basis vectors is somewhat arbitrary, one can ask what functions of matrix elements $A_{i,j}$ are invariant under the rescaling of the two basis sets. There are $n^2-2n+1=(n-1)^2$ such invariants ($1$ comes from the fact that rescaling by $\lambda 1\!\!1$ has the same effect on both sides). Such invariants can be constructed via $\tilde A_{i,j}=A_{i+1,j}A_{i,j+1}/(A_{i,j}A_{i+1,j+1})$. This scale-invariant matrix has a remarkably simple form {\tiny$\left(
\begin{array}{cc}
 -1 & 0 \\
 1 & -1 \\
\end{array}
\right)$}, which is the same for all examples below with $s_{1,2,3}>0$. }
\begin{align}
\begin{pmatrix}
    \langle J_1 J_1 J_1 \rangle_{\text{F.B.}}\\
    \langle J_1 J_1 J_1 \rangle_{\text{F.F.}}\\
    \langle J_1 J_1 J_1 \rangle_{\text{odd}}
    \end{pmatrix}&=\frac{8}{\pi ^3}\left(
\begin{array}{ccc}
 -2 & 1 & 1 \\
 -2 & -1 & -1 \\
 0 &  \pi i & - \pi i  \\
\end{array}
\right)\begin{pmatrix}
    \VV^{(\pm1,\mp1,\mp1)}\\
    \VVV^{+1,+1,+1}\\
    \VVV^{-1,-1,-1}
    \end{pmatrix}\,,
\end{align}
which clearly shows that the pure Yang-Mills vertex leads to the equal number of bosons and fermions. On the other hand $\VVV^{+1,+1,+1}+\VVV^{-1,-1,-1}$ is parity even and results in the difference $X=\langle J_1 J_1 J_1 \rangle_{\text{F.B.}}-\langle J_1 J_1 J_1 \rangle_{\text{F.F.}}$. Since these two vertices are abelian, i.e. trivially gauge invariant, $X$ satisfies the trivial Ward identities. The bulk difference $\VVV^{+1,+1,+1}-\VVV^{-1,-1,-1}$ is parity odd, i.e. leads to $\langle J_1 J_1 J_1 \rangle_{\text{odd}}$, which again satisfies the trivial Ward identity. We also note that the topological Pontryagin term does not induce anything.

%%%%%%%%%%%%%%%%%%%%%%%%%%%%%%%%%%%%%%%%%%%%%%%%%%%%%%%%%%%%%%%%%%%%%%%
\subsection{Gravity with higher derivative corrections}
\label{subsec:GR}
%%%%%%%%%%%%%%%%%%%%%%%%%%%%%%%%%%%%%%%%%%%%%%%%%%%%%%%%%%%%%%%%%%%%%%%
We begin with the usual Einstein-Hilbert action with the cosmological constant, but written in MacDowell-Mansouri form \cite{MacDowell:1977jt}:
\begin{align}
    S_{\text{EH}}&= \tfrac{1}4\int \mathcal{R}_{\ga\ga}\wedge \mathcal{R}^{\ga\ga} -\mathcal{R}_{\gad\gad}\wedge \mathcal{R}^{\gad\gad} \,.
\end{align}
Here, $\mathcal{R}^{\ga\ga}\equiv R^{\ga\ga}-e\fud{\ga}{\gdd}\wedge e^{\ga\gdd}$, where $R^{\ga\ga}\equiv d\omega^{\ga\ga} -\omega\fud{\ga}{\gb}\wedge \omega^{\ga\gb}$ is the Riemann two-form provided spin-connection $\omega^{\ga\gb}$ satisfies the Torsion constraint, idem. for $\mathcal{R}^{\gad\gad}$. Vierbein $e^{\ga\gad}$ and (anti)-self-dual components  $\omega^{\ga\gb}$, $\omega^{\gad\gbd}$ of the spin-connection are considered as independent fields in the action. The equations of motion imply the Torsion constraint
\begin{align}
    de^{\ga\gad}&= \omega\fud{\ga}{\gb} \wedge e^{\gb\gad} +\omega\fud{\gad}{\gbd}\wedge e^{\ga\gbd} \,,
\end{align}
and Einstein equations, which can be extracted either from $\mathcal{R}^{\ga\ga}$ or $\mathcal{R}^{\gad\gad}$. We choose $\mathcal{R}^{\ga\ga}$ and it is useful to write the equations as 
\begin{align}
    \mathcal{R}^{\ga\ga}&= E_{\gb\gb} W^{\ga\ga\gb\gb}\,.
\end{align}
We recall that $E^{\ga\ga}\equiv e\fud{\ga}{\gdd}\wedge e^{\ga \gdd}$. Here $W^{\ga\gb\gc\gd}$ parameterizes the self-dual component of the Weyl tensor. For free gravitons we define the fluctuating vierbein $e^{\ga\gad}$ (as opposite to the background veirbein $h^{\ga\gad}$) and call fluctuating spin-connections $\omega^{\ga\gb}$, $\omega^{\gad \gbd}$ by the same names. The free equations read
\begin{align}
    \nabla e^{\ga\gad}&= h\fud{\ga}{\gdd}\wedge \omega^{\gad\gdd}+h\fdu{\gc}{\gad}\wedge \omega^{\ga\gc}\,, & 
    \begin{aligned}
            \nabla \omega^{\ga\gb}&= 2h\fud{(\ga}{\gdd}\wedge e^{\gb)\gdd}+H_{\gc\gd} W^{\ga\gb\gc\gd}\,, \\
    \nabla \omega^{\gad\gbd}&= 2h\fdu{\gc}{(\gad}\wedge e^{\gc\gbd)} +H_{\gdd\gdl} W^{\gad\gbd\gdd\gdl}\,,
    \end{aligned}
\end{align}
where $\nabla$ is the Lorentz covariant derivative, see Appendix \ref{app:notation} for more detail on our notation. The boundary-to-bulk propagator \eqref{propagator} for the spin-$s$ field immediately tells us that we can define $e^{\ga\gad}\equiv h_{\gb\gbd}\Phi^{\ga\gb,\gad\gbd}=K h^{\gb\gbd}\xi_\gb \brxi_\gbd \xi^\ga \brxi^{\gad}$. Likewise, from the equations of motion we find 
\begin{align}
     \omega^{\ga\ga}&= -2K h^{\gb\gbd} \xi_\gb \brxi_\gbd \xi^\ga \xi^{\ga}\,, & 
     W_{\ga(4)}&= -3K\xi_{\ga(4)} \,,
\end{align}
and similarly for the conjugates thereof. 
Again, there is a topological invariant 
\begin{align}
    S_{\text{top}}&= \tfrac{1}4\int \mathcal{R}_{\ga\ga} \mathcal{R}^{\ga\ga} +\mathcal{R}_{\gad\gad} \mathcal{R}^{\gad\gad} 
\end{align}
that can be used to eliminate any one of the two terms in the action. 
Now, the cubic vertex can be extracted as follows\footnote{We use the fact that the MacDowell-Mansouri action is the Einstein action with all the necessary holographic counterterms added \cite{Miskovic:2009bm}. In particular, its on-shell value is given by the Weyl gravity action \cite{Miskovic:2009bm}. }
\begin{align}
    \tfrac12\int \left(\mathcal{R}_{\ga\ga} \mathcal{R}^{\ga\ga}\right)\big|_{3}&= \int \left(\mathcal{R}_{\ga\ga}\right)\big|_{1} \left(\mathcal{R}^{\ga\ga}\right)\big|_{2}=
    -\int  H^{\gb\gb} W_{\ga\ga\gb\gb} \wedge (\omega\fud{\ga}{\gc}\wedge \omega^{\ga\gc}+e\fud{\ga}{\gdd}\wedge e^{\ga\gdd}) \,,
\end{align}
where ${}|_{i}$ means the order in the fluctuating fields and in the last step we used the free equations of motion to replace $(\mathcal{R}_{\ga\ga})|_1$ with its free field value $H^{\gb\gb} W_{\ga\ga\gb\gb}$. Now, we plug in the on-shell values for all these fields to get
\begin{align}
   \tfrac12\int \left(\mathcal{R}_{\ga\ga} \mathcal{R}^{\ga\ga}\right)\big|_{3}&=\tfrac12\VV^{\pm2,\pm2,\mp2}\,,
\end{align}
where the amplitude is
\begin{align}\label{spintwoppm}
   \VV^{\pm2,\pm2,\mp2}&= \int \vol K_1K_2 K_3 (\xi_1\xi_2)^2(\xi_3\xi_1)^2 (\bar \xi_2 \bar \xi_3)\left[4 (\xi_2\xi_3) +(\bar \xi_2 \bar \xi_3)\right]\,.
\end{align}
This corresponds to the sum of $\VVV^{-2,+2,+2}$ and $\VVV^{+2,-2,-2}$ of the Einstein-Hilbert vertex. The second half arises from $-\mathcal{R}_{\gad\gad} \mathcal{R}^{\gad\gad}$ and results in the same amplitude, whose actual expression is given by the complex conjugate of the $\VV^{\pm2,\pm2,\mp2}$ with plus sign (note that $H_{\gad\gad}\wedge H^{\gad\gad}=-H_{\ga\ga}\wedge H^{\ga\ga}$, see Appendix \ref{app:notation}). Finally,
\begin{align}
      S_{EH}\Big|_{3}&= \tfrac1{4} \left(\VV^{\pm2,\pm2,\mp2} +\VV^{\mp2,\mp2,\pm2}\right)=\tfrac12\VV^{\pm2,\pm2,\mp2}\,.
\end{align}
As in the Yang-Mills case there are two higher derivative vertices. A generic linear combination of them violates parity and they are also needed to get non-supersymmetric correlation functions. We parameterize them as\footnote{The strange prefactor is due to the normalization of the Weyl tensor, which is $-3$.}
\begin{align}
    -\tfrac{1}{27}a_2 \int \vol W\fdu{\ga\ga}{\gb\gb} W\fdu{\gb\gb}{\gc\gc} W\fdu{\gc\gc}{\ga\ga}-\tfrac{1}{27}a_3 \int \vol W\fdu{\gad\gad}{\gbd\gbd} W\fdu{\gbd\gbd}{\gdd\gdd} W\fdu{\gdd\gdd}{\gad\gad} \,.
\end{align}
Their contribution to the amplitude is obvious:
\begin{align}
    a_2 \VVV^{+2,+2,+2} + a_3 \VVV^{-2,-2,-2} \,,
\end{align}
where the elementary amplitudes are
\begin{align}
    \VVV^{+2,+2,+2}&=\int \vol K_1K_2K_3 (\xi_1 \xi_2)^2 (\xi_2 \xi_3)^2(\xi_3 \xi_1)^2\,,\\
    \VVV^{-2,-2,-2}&=\int \vol K_1K_2K_3 (\brxi_1 \brxi_2)^2 (\brxi_2 \brxi_3)^2(\brxi_3 \brxi_1)^2\,,
\end{align}
and the separation between different helicity structures is possible. 
The basis of the CFT correlators consists of 
{\footnotesize
\begin{align*}
    \langle J_2 J_2 J_2 \rangle_{\text{F.B.}} &=-\frac{5}{32} P_1 P_2 P_3 Q_1 Q_2 Q_3-\frac{5}{192} \left(P_1^4 Q_1^2+P_2^4 Q_2^2+P_3^4 Q_3^2\right)+\frac{1}{4} P_1^2 P_2^2 P_3^2+\frac{25}{512} Q_1^2 Q_2^2 Q_3^2\,, \\
    \langle J_2 J_2 J_2 \rangle_{\text{F.F.}} &=-\frac{1}{24} P_1 P_2 P_3 \left(5 Q_1 Q_2 Q_3-4 P_1 P_2 P_3\right)\,,\\
    \langle J_2 J_2 J_2 \rangle_{\text{odd}} &= \left(P_1 P_2 P_3+Q_1 Q_2 Q_3\right) \left(P_2 P_3 S_1+P_1 P_3 S_2+P_1 P_2 S_3\right)\,,
\end{align*}}%
where we slightly simplified the free boson/fermion structures and made a particular choice for the odd structure (it is unique up to the identities). The final result can be written as $3\times 3$ matrix:
\begin{align}
    \begin{pmatrix}
    \VV^{(\pm2,\mp2,\mp2)}\\
    \VVV^{+2,+2,+2}\\
    \VVV^{-2,-2,-2}
    \end{pmatrix}&= \frac{9 \pi ^2}{512}
\left(
\begin{array}{ccc}
 \pi  & \pi  & 0 \\
 \frac{5 \pi }{3} & -\frac{5 \pi }{3}  & +\frac{10 i}{27} \\
 \frac{5 \pi }{3} & -\frac{5 \pi }{3}  & -\frac{10 i}{27} \\
\end{array}
\right)\begin{pmatrix}
    \langle J_2 J_2 J_2 \rangle_{\text{F.B.}}\\
    \langle J_2 J_2 J_2 \rangle_{\text{F.F.}}\\
    \langle J_2 J_2 J_2 \rangle_{\text{odd}}
    \end{pmatrix}\,.
\end{align}
The inverse transformation  is
\begin{align}
   \begin{pmatrix}
    \langle J_2 J_2 J_2 \rangle_{\text{F.B.}}\\
    \langle J_2 J_2 J_2 \rangle_{\text{F.F.}}\\
    \langle J_2 J_2 J_2 \rangle_{\text{odd}}
    \end{pmatrix}&=\frac{384}{5 \pi ^3} \left(
\begin{array}{ccc}
 \frac{10}{27} & \frac{1}{9} & \frac{1}{9} \\
 \frac{10}{27} & -\frac{1}{9} & -\frac{1}{9} \\
 0 & - \pi i  &  \pi i  \\
\end{array}
\right) \begin{pmatrix}
    \VV^{(\pm2,\mp2,\mp2)}\\
    \VVV^{+2,+2,+2}\\
    \VVV^{-2,-2,-2}
    \end{pmatrix}\,.
\end{align}
Everything said about the spin-one self-interactions is valid for the spin-two case as well.

%%%%%%%%%%%%%%%%%%%%%%%%%%%%%%%%%%%%%%%%%%%%%%%%%%%%%%%%%%%%%%%%%%%%%%%
\subsection{Mixed vertices}
%%%%%%%%%%%%%%%%%%%%%%%%%%%%%%%%%%%%%%%%%%%%%%%%%%%%%%%%%%%%%%%%%%%%%%%
As to have more examples we also consider some mixed low-spin vertices below.   

\paragraph{$\boldsymbol{2-1-1}$. } 
These are interactions between two Yang-Mills fields and the graviton. There are two types of vertices: the usual gravitation coupling of the spin-one field's stress-tensor to metric and the higher derivative one. In the two-component spinor language the stress-tensor is $T_{\ga\ga,\gad\gad}=F_{\ga\ga} F_{\gad\gad}$. We can start with the same action \eqref{YMaction}
\begin{align}\label{YMactionA}
    S_{\text{YM}}&=\tfrac12 \Tr \int E_{\ga\ga}\wedge E^{\ga\ga} \left[ F_{\ga\ga}F^{\ga\ga}+F_{\gad\gad}F^{\gad\gad} \right]\,,
\end{align}
but take into account that the volume form is built from $E^{\ga\ga}\equiv e\fud{\ga}{\gdd}\wedge e^{\ga \gdd}$ where $e^{\ga\gad}$ is the full vierbein rather than its AdS${}_4$ value $h^{\ga\gad}$. The two-derivative vertex is obtained by varying the action once with respect to $e^{\ga\gad}$ and twice with respect to $A$.\footnote{Note that $F$'s definition, \eqref{atoPsi}, depends on $e^{\ga\gad}$. Therefore, when taking the variation with respect to $e^{\ga\gad}$ we need to take into account both explicit dependence on $e^{\ga\gad}$ and the implicit one through $F$.} As a result we find 
\begin{align}
    -8\int e^{\ga\gad} \hat{h}^{\ga\gad} F_{\ga\ga} F_{\gad\gad}&=\tfrac12\VV^{\pm2,\pm1,\mp1}\,.
\end{align}
This canonical gravitational coupling leads to the following amplitude\footnote{If we subtract the complex conjugate below, i.e. anti-symmetrize over $(23)$, we get a structure that is odd, conserved with respect to both spin-one legs and is not conserved for the spin-two leg.} 
\begin{align}
    \VV^{\pm2,\pm1,\mp1}&=\int \vol K_1 K_2 K_3 (\xi_1 \xi_2)^2 (\brxi_3 \brxi_1)^2+\text{c.c.}\,,
\end{align}
where we have added $\text{c.c.}$ since the stress-tensor is diagonal in the spin-one fields (in case we have several) and, hence, should be $2\leftrightarrow3$ symmetric. This amplitude is not pure and is equal to $\VVV^{+2,+1,-1}+\VVV^{-2,+1,-1}$. We will also need the higher derivative couplings 
\begin{align}
    \int \vol W^{\ga\ga\ga\ga} F_{\ga\ga} F_{\ga\ga}&=-\tfrac34\VVV^{+2,+1,+1}\,,  && \int \vol W^{\gad\gad\gad\gad} F_{\gad\gad} F_{\gad\gad}=-\tfrac34\VVV^{-2,-1,-1}\,,
\end{align}
which leads to the following amplitudes 
\begin{align}
    \VVV^{+2,+1,+1}&= \int \vol K_1 K_2 K_3 (\xi_1 \xi_2)^2 (\xi_3 \xi_1)^2\,, & \VVV^{-2,-1,-1}&=\int \vol K_1 K_2 K_3 (\brxi_1 \brxi_2)^2 (\brxi_3 \brxi_1)^2\,.
\end{align}
The basis to expand is chosen as 
\besubeqs
\begin{align}
    \langle J_2 J_1 J_1 \rangle_{\text{F.B.}} &= \frac{1}{32} \left(4 P_2^2 Q_1 Q_2+2 P_1^2 Q_1^2+4 P_3^2 Q_1 Q_3+8 P_3^2 P_2^2+Q_1^2 Q_2 Q_3\right)\,, \\
    \langle J_2 J_1 J_1 \rangle_{\text{F.F.}} &=-\frac{1}{2} P_1 P_2 P_3 Q_1\,,\\
    \langle J_2 J_1 J_1 \rangle_{\text{odd}} &= Q_1 \left(P_2 Q_2 S_2+P_3 Q_3 S_3\right)\,.
\end{align}
\esubeqs
The final result can be written as 
\begin{align}
    \begin{pmatrix}
    \VV^{\pm2,\pm1,\mp1}\\
    \VVV^{+2,+1,+1}\\
    \VVV^{-2,-1,-1}
    \end{pmatrix}&= \frac{\pi ^2}{128}
\left(
\begin{array}{ccc}
 6 \pi  & 6 \pi  & 0 \\
 3 \pi  & -3 \pi  & 2 i \\
 3 \pi  & -3 \pi  & -2 i \\
\end{array}
\right)\begin{pmatrix}
    \langle J_2 J_1 J_1 \rangle_{\text{F.B.}}\\
    \langle J_2 J_1 J_1 \rangle_{\text{F.F.}}\\
    \langle J_2 J_1 J_1 \rangle_{\text{odd}}
    \end{pmatrix}\,.
\end{align}
The scale invariant matrix is the same as before. The inverse transformation reads
\begin{align}
\begin{pmatrix}
    \langle J_2 J_1 J_1 \rangle_{\text{F.B.}}\\
    \langle J_2 J_1 J_1 \rangle_{\text{F.F.}}\\
    \langle J_2 J_1 J_1 \rangle_{\text{odd}}
    \end{pmatrix}&=\frac{32}{3 \pi ^3}\left(
\begin{array}{ccc}
 1 & 1 & 1 \\
 1 & -1 & -1 \\
 0 & -3  \pi i  & 3 \pi i  \\
\end{array}
\right)\begin{pmatrix}
    \VV^{\pm2,\pm1,\mp1}\\
    \VVV^{+2,+1,+1}\\
    \VVV^{-2,-1,-1}
    \end{pmatrix}\,.
\end{align}

\paragraph{$\boldsymbol{2-2-1}$. } Next, we consider all possible vertices between a Yang-Mills field and two massless spin-two fields that can be reached. The latter have to be charged for the vertex to exist \cite{Metsaev:2005ar} and, hence, the particles cannot be the graviton \cite{Boulanger:2000rq}.\footnote{Note that such vertices do not appear in any of the theories we are aware of, except for the hypothetical higher spin gravity since they are required to reproduce the right three-point functions. } According to the classification there are three vertices: $\VVV^{+2,+2,+1}$ with $5$ derivatives, $\VVV^{+2,+2,-1}$ with $3$ derivatives and $\VVV^{+2,-2,+1}$ with $1$ derivative. The latter vertex is present in the higher spin extension \cite{Krasnov:2021nsq} of SDYM and cannot be written in terms of $\Phi_{\ga(s),\gad(s)}$, see e.g. \cite{Metsaev:2005ar, Conde:2016izb}. We cannot separate $1$- and $3$-derivative vertices. A specific linear combination can be written as
\begin{align}
    \Tr \int  F^{\ga\ga} H_{\ga\ga} \wedge\left( \{\omega_{\gb\gb},\omega^{\gb\gb} \}+\{\omega_{\gbd\gbd},\omega^{\gbd\gbd} \}+2\{e_{\gb\gbd},e^{\gb\gbd} \}\right)+ 2\{ A,  \omega_{\ga\ga}\} \wedge W^{\ga\ga\ga\ga} H_{\ga\ga} \,,
\end{align}
where in the last term the anti-symmetrization over the two `gravitons' is understood (with the usual normalization $1/2$). The corresponding amplitude reads
\begin{align}
\begin{aligned}
&\VV^{\pm2,\pm2,\mp1}=\text{Tr}\int \text{vol }K_{1}K_{2}K_{3}\big[3\xisc{1}{2}^3 \xisc{2}{3}\xiscb{3}{1}+3\xisc{1}{2}^3\xisc{3}{1}\xiscb{2}{3}+\\&\quad+\xisc{1}{2}^2\xiscb{1}{2}\xisc{2}{3}\xisc{3}{1}+\tfrac12\xisc{1}{2}\xiscb{1}{2}^2\xisc{2}{3}\xisc{3}{1}+\xiscb{1}{2}^3\xisc{2}{3}\xisc{3}{1}\big]\,.
\end{aligned}
\end{align}
The $5$-derivative vertices are the usual abelian ones:
\begin{align}
    \Tr\int \vol F^{\ga\ga} W_{\ga\gb\gb\gb}W\fdu{\ga}{\gb\gb\gb}&= \tfrac{9}{2} \VVV^{+2,+2,+1}\,,
\end{align}
and its complex conjugate. The amplitude is
\begin{align}
\VVV^{+2,+2,+1}&=\Tr\int \vol K_{1}K_{2}K_{3}\left( \xi _{1}\xi
_{2}\right)^3 \left( \xi _{3}\xi _{1}\right) \left( \xi _{2}\xi _{3}\right)\,.
\end{align}%
The basis to expand is chosen as 
{\footnotesize\begin{align*}
    \langle J_2 J_2 J_1 \rangle_{\text{F.B.}} &= \frac{1}{384} \big(12 P_1^2 Q_1 \left(4 P_3^2+Q_1 Q_2\right)+12 P_2^2 Q_2 \left(4 P_3^2+Q_1 Q_2\big)+Q_3 \left(24 P_3^2 Q_1 Q_2+8 P_3^4+3 Q_1^2 Q_2^2\right)\right)\,, \\
    \langle J_2 J_2 J_1 \rangle_{\text{F.F.}} &=\frac{1}{36} P_1 P_2 P_3 \left(-6 P_3^2-9 Q_1 Q_2\right)\,,\\
    \langle J_2 J_2 J_1 \rangle_{\text{odd}} &= \frac{1}{6} \left(2 P_3^3 Q_3 S_3+3 P_3 Q_1 Q_2 Q_3 S_3+P_1 Q_1^2 Q_2 S_1+P_2 Q_1 Q_2^2 S_2+4 P_1 P_2 P_3^2 S_3\right)\,.
\end{align*}}%
The final result can be written as 
\begin{align}
    \begin{pmatrix}
    \VV^{\pm2,\pm2,\mp1}\\
    \VVV^{+2,+2,+1}\\
    \VVV^{-2,-2,-1}
    \end{pmatrix}&= \frac{\pi ^2}{256}
\left(
\begin{array}{ccc}
 -9 \pi  & -9 \pi  & 0 \\
 6 \pi  & -6 \pi  & 4 i \\
 6 \pi  & -6 \pi  & -4 i \\
\end{array}
\right)\begin{pmatrix}
    \langle J_2 J_2 J_1 \rangle_{\text{F.B.}}\\
    \langle J_2 J_2 J_1 \rangle_{\text{F.F.}}\\
    \langle J_2 J_2 J_1 \rangle_{\text{odd}}
    \end{pmatrix}\,,
\end{align}
The inverse transformation reads
\begin{align}
\begin{pmatrix}
    \langle J_2 J_2 J_1 \rangle_{\text{F.B.}}\\
    \langle J_2 J_2 J_1 \rangle_{\text{F.F.}}\\
    \langle J_2 J_2 J_1 \rangle_{\text{odd}}
    \end{pmatrix}&=\frac{32}{9 \pi ^3}\left(
\begin{array}{ccc}
 -4 & 3 & 3 \\
 -4 & -3 & -3 \\
 0 & -9  \pi i  & 9  \pi i  \\
\end{array}
\right)\begin{pmatrix}
    \VV^{\pm2,\pm2,\mp1}\\
    \VVV^{+1,+1,+1}\\
    \VVV^{-1,-1,-1}
    \end{pmatrix}\,.
\end{align}

\paragraph{$\boldsymbol{0-1-1}$.} Let us also consider several examples involving one scalar field. There should be two structures, which are easily seen to be 
\begin{align}
    &4\int \vol \phi F_{\ga\ga}F^{\ga\ga}\,, && 4\int \vol \phi F_{\gad\gad}F^{\gad\gad}\,,
\end{align}
and the corresponding amplitudes are
\begin{align}
    \VVV^{0,+1,+1}&= \int \vol K_1^\Delta K_2K_3 (\xi_2 \xi_3)^2\,, & \VVV^{0,-1,-1}&= \int \vol K_1^\Delta K_2K_3 (\brxi_2 \brxi_3)^2 \,,
\end{align}
where $\Delta=1$ or $\Delta=2$. In the first case we have
\begin{align}
    \langle J_0 J_1 J_1\rangle_{\text{F.B.}}&=\frac{1}{4} \left(2 P_1^2+Q_2 Q_3\right)\,, &
    \langle J_0 J_1 J_1\rangle_{\text{C.F.}}&=\frac{4 P_1 S_1}{\pi }\,,
\end{align}
where C.F.\ means critical fermion model. The dictionary is
\begin{align}
    \begin{pmatrix}
    \VVV^{0,+1,+1}\\
    \VVV^{0,-1,-1}
    \end{pmatrix}&=  \frac{\pi ^3}{32}
\left(
\begin{array}{cc}
 2 & i \\
 2 & -i \\
\end{array}
\right)\begin{pmatrix}
    \langle J_0 J_1 J_1\rangle_{\text{F.B.}}\\
    \langle J_0 J_1 J_1\rangle_{\text{C.F.}}
    \end{pmatrix}\,,
\end{align}
and the scale-invariant, which is just a number, is trivially $-1$. For the case $\Delta=2$ we have 
\begin{align}
    \langle \tilde J_0 J_1 J_1\rangle_{\text{F.F.}}&=i P_1 S_1\,, &
    \langle \tilde J_0 J_1 J_1\rangle_{\text{C.B.}}&=\frac{2 Q_2 Q_3}{\pi }\,,
\end{align}
where C.B.\ means critical vector model (critical boson). The dictionary is
\begin{align}
    \begin{pmatrix}
    \VVV^{0,+1,+1}\\
    \VVV^{0,-1,-1}
    \end{pmatrix}&=  \frac{\pi ^3}{32}
\left(
\begin{array}{cc}
 1 & 1 \\
 -1 & 1 \\
\end{array}
\right)\begin{pmatrix}
    \langle \tilde J_0 J_1 J_1\rangle_{\text{F.F.}}\\
    \langle \tilde J_0 J_1 J_1\rangle_{\text{C.B.}}
    \end{pmatrix}\,.
\end{align}

\paragraph{$\boldsymbol{0-2-2}$.} Likewise, there should be two structures:
\begin{align}
    &\tfrac19\int \vol \phi W_{\ga\ga\ga\ga}W^{\ga\ga\ga\ga}\,, && \tfrac19\int \vol \phi W_{\gad\gad\gad\gad}W^{\gad\gad\gad\gad}\,,
\end{align}
and the corresponding amplitudes are
\begin{align}
    \VVV^{0,+2,+2}&= \int \vol K_1^\Delta K_2K_3 (\xi_2 \xi_3)^4\,, & \VVV^{0,-2,-2}&= \int \vol K_1^\Delta K_2K_3 (\brxi_2 \brxi_3)^4 \,,
\end{align}
where $\Delta=1,2$. For $\Delta=1$ we have
\begin{align}
    \langle J_0 J_2 J_2\rangle_{\text{F.B.}}&=\frac{1}{24} \left(3 P_1^2 Q_2 Q_3+P_1^4+\frac{3}{8} Q_2^2 Q_3^2\right)\,,\\
    \langle J_0 J_2 J_2\rangle_{\text{C.F.}}&=\frac{4 S_1 \left(P_1 Q_2 Q_3+2 P_1^3\right)}{9 \pi }\,,
\end{align}
and the dictionary is
\begin{align}
    \begin{pmatrix}
    \VVV^{0,+2,+2}\\
    \VVV^{0,-2,-2}
    \end{pmatrix}&= \frac{3 \pi ^3}{256}
\left(
\begin{array}{cc}
 2 & i \\
 2 & -i \\
\end{array}
\right)\begin{pmatrix}
    \langle J_0 J_2 J_2\rangle_{\text{F.B.}}\\
    \langle J_0 J_2 J_2\rangle_{\text{C.F.}}
    \end{pmatrix}\,.
\end{align}
For $\Delta=2$ we have
\besubeqs
\begin{align}
    \langle \tilde J_0 J_2 J_2\rangle_{\text{F.F.}}&=\frac{1}{6} i P_1 S_1 \left(P_1^2+\frac{3 Q_2 Q_3}{2}\right)\,,\\
    \langle \tilde J_0 J_2 J_2\rangle_{\text{C.B.}}&=\frac{8 P_1^2 Q_2 Q_3-4 P_1^4+2 Q_2^2 Q_3^2}{9 \pi }\,,
\end{align}
\esubeqs
\begin{align}
    \begin{pmatrix}
    \VVV^{0,+2,+2}\\
    \VVV^{0,-2,-2}
    \end{pmatrix}&= \frac{3 \pi ^3}{256}
\left(
\begin{array}{cc}
 1 & 1 \\
 -1 & 1 \\
\end{array}
\right)\begin{pmatrix}
    \langle \tilde J_0 J_2 J_2\rangle_{\text{F.F.}}\\
    \langle \tilde J_0 J_2 J_2\rangle_{\text{C.B.}}
    \end{pmatrix}\,.
\end{align}
Correlation functions of type $0-s_1-s_2$ were studied in \cite{Giombi:2016zwa,Sezgin:2017jgm}.

\paragraph{$\boldsymbol{0-1-2}$.} This is another subtle example without higher spin fields (in addition to $1-2-2$). The amplitude/light-cone/spinor-helicity analysis offers us  $\VVV^{0,+1,+2}$, $\VVV^{0,-1,+2}$ and complex conjugates thereof. One cannot write the vertex with one derivative, $\VVV^{0,-1,+2}$, by invoking symmetric covariant fields $\Phi_{\ga(s),\gad(s)}$ or by extrapolating the light-cone results from $d>4$ to $d=4$ \cite{Metsaev:2005ar, Conde:2016izb}. Instead, one can write down two `$3$-derivative' vertices as
\begin{align}
    \int  \Phi(\nabla^{\ga\gad}F^{\ga\ga})(\Phi_{\ga\ga\ga,\gad})\oplus \text{c.c.}\,,
\end{align}
where we recall that $\Phi_{\ga\ga\ga,\gad}=\nabla\fdu{\ga}{\gbd}\Phi_{\ga\ga,\gad\gbd}$ according to \eqref{allmycurls}. The vertex is manifestly gauge-invariant with respect to spin-one and, hence, is conserved on this leg. It is, however, not conserved with respect to spin-two, as the CFT analysis also shows \cite{Giombi:2011rz}. 
They lead to two amplitudes
\begin{align}
    \VV^{0,+1,+2}&= \int \vol K_1^\Delta K_2K_3(\xi_2\xi_3)^2 (\xi_3 F_2 \brxi_3)\,, &
    \VV^{0,-1,-2}&= \int \vol K_1^\Delta K_2K_3(\brxi_2\brxi_3)^2 (\xi_3 F_2 \brxi_3)\,.
\end{align}
For $\Delta=1$ we have
\begin{align}
    \langle J_0 J_1 J_2\rangle_{\text{F.B.}}&=-\frac{1}{16} \left(4 P_1^2 Q_3 + Q_2 Q_3^2 \right)\,,&
    \langle J_0 J_1 J_2\rangle_{\text{C.F.}}&=-\frac{4 i P_1 Q_3 S_1}{3 \pi }\,,
\end{align}
\begin{align}
    \begin{pmatrix}
    \VV^{0,+1,+2}\\
    \VV^{0,-1,-2}
    \end{pmatrix}&= \frac{\pi ^3}{64}
\left(
\begin{array}{cc}
 2 & 1 \\
 2 & -1 \\
\end{array}
\right)\begin{pmatrix}
    \langle J_0 J_1 J_2\rangle_{\text{F.B.}}\\
    \langle J_0 J_1 J_2\rangle_{\text{C.F.}}
    \end{pmatrix}\,.
\end{align}
For $\Delta=2$ we find
\begin{align}
    \langle \tilde J_0 J_1 J_2\rangle_{\text{F.F.}}&=-\frac{1}{2} i P_1 Q_3 S_1\,, &
    \langle \tilde J_0 J_1 J_2\rangle_{\text{C.B.}}&=-\frac{2 Q_3 \left(P_1^2+Q_2 Q_3\right)}{3 \pi }\,,
\end{align}
\begin{align}
    \begin{pmatrix}
    \VV^{0,+1,+2}\\
    \VV^{0,-1,-2}
    \end{pmatrix}&= \frac{\pi ^3}{64}
\left(
\begin{array}{cc}
 1 & 1 \\
 -1 & 1 \\
\end{array}
\right)\begin{pmatrix}
    \langle \tilde J_0 J_1 J_2\rangle_{\text{F.F.}}\\
    \langle \tilde J_0 J_1 J_2\rangle_{\text{C.B.}}
    \end{pmatrix}\,.
\end{align}

%%%%%%%%%%%%%%%%%%%%%%%%%%%%%%%%%%%%%%%%%%%%%%%%%%%%%%%%%%%%%%%%%%%%%%%
\subsection{Higher spins}\label{subsec:HS}
%%%%%%%%%%%%%%%%%%%%%%%%%%%%%%%%%%%%%%%%%%%%%%%%%%%%%%%%%%%%%%%%%%%%%%%
It is hard to explore correlation functions for spins outside the triangle as long as we do not go beyond spin-two: $0-0-1$, $0-0-2$ have a very poor structure. One exception is $0-1-2$, which was studied above. Here we also consider the example of $1-1-3$, which is $\langle J_1 J_1 J_3\rangle$, see also a detailed discussion on the CFT side in \cite{Giombi:2016zwa}. Then, we generalize and summarize the results for all spins. 

\paragraph{$\boldsymbol{1-1-3}.$} In principle, one has $\VVV^{+1,+1,+3}$, $\VVV^{+1,-1,+3}$, $\VVV^{-1,-1,+3}$ at disposal and the complex conjugates thereof. The lowest-derivative one, $\VVV^{-1,-1,+3}$, seems unreachable, but a linear combination of $\VVV^{-1,-1,+3}$ with $\VVV^{+1,-1,+3}$ plus the complex conjugate is easily found to be
\begin{align}
    \VV^{\pm 1,\mp1,\pm3}&=\int \vol \left[\xisc{3}{1}^2 \xiscb{2}{3}^2 (\xi_3 F_1\brxi_3)+\xiscb{3}{1}^2 \xisc{2}{3}^2 (\xi_3 F_1\brxi_3)\right] \,.
\end{align}
The highest-derivative vertices cannot be made pure as well: they are manifestly gauge-invariant with respect to spin-one, but not for spin-three:
\begin{align}
    \VV^{+1,+1,+3}&=\int \vol  \xisc{3}{1}^2 \xisc{2}{3}^2 (\xi_3 F_1\brxi_3) \,, \\
    \VV^{-1,-1,-3}&=\int \vol  \xiscb{3}{1}^2 \xiscb{2}{3}^2 (\xi_3 F_1\brxi_3) \,,
\end{align}
where we chose to have $F_1$, but $F_2$ would give the same expression via integration by parts. The basis of conformal structures is chosen as
{\footnotesize\begin{align*}
    \langle J_1 J_1 J_3 \rangle_{\text{F.B.}} &= \frac{1}{192} Q_3 \left(6 P_1^2 \left(4 P_2^2+Q_1 Q_3\right)+Q_3 \left(6 P_2^2 Q_2+Q_3 \left(2 P_3^2+Q_1 Q_2\right)\right)\right)\,, \\
    \langle J_1 J_1 J_3 \rangle_{\text{F.F.}} &=-\frac{1}{8} P_1 P_2 P_3 Q_3^2\,,\\
    \langle J_1 J_1 J_3 \rangle_{\text{odd}} &= 2 Q_3^2 \left(P_1 Q_1 S_1+P_2 Q_2 S_2+P_1 P_2 S_3\right)\,.
\end{align*}}%
The final result can be written as
\begin{align}
    \begin{pmatrix}
    \VV^{\pm1,\mp1,\pm3}\\
    \VV^{+1,+1,+3}\\
    \VV^{-1,-1,-3}
    \end{pmatrix}&= \frac{5 \pi ^2}{128}
\left(
\begin{array}{ccc}
 2 \pi  & 2 \pi  & 0 \\
 \pi  & -\pi  & \frac{i}{15} \\
 \pi  & -\pi  & -\frac{i}{15} \\
\end{array}
\right)\begin{pmatrix}
    \langle J_1 J_1 J_3 \rangle_{\text{F.B.}}\\
    \langle J_1 J_1 J_3 \rangle_{\text{F.F.}}\\
    \langle J_1 J_1 J_3 \rangle_{\text{odd}}
    \end{pmatrix}\,.
\end{align}
It may also be convenient to choose the basis of interactions that have definite parity, for which we find
\begin{align}
    \begin{pmatrix}
    \VV^{\pm1,\mp1,\pm3}\\
    \VV^{+1,+1,+3}+\VV^{-1,-1,-3}\\
    \VV^{+1,+1,+3}-\VV^{-1,-1,-3}
    \end{pmatrix}&= \frac{5 \pi ^3}{64}
\left(
\begin{array}{ccc}
 1 & 1 & 0 \\
 1 & -1 & 0 \\
 0 & 0 & \frac{i}{15 \pi } \\
\end{array}
\right)\begin{pmatrix}
    \langle J_1 J_1 J_3 \rangle_{\text{F.B.}}\\
    \langle J_1 J_1 J_3 \rangle_{\text{F.F.}}\\
    \langle J_1 J_1 J_3 \rangle_{\text{odd}}
    \end{pmatrix}\,.
\end{align}

\paragraph{All spins.} Several simple facts can be pointed out for the most general cases. Suppose that we are interested in $\langle J_{s_1} J_{s_2} J_{s_3} \rangle$ for generic spins, $s_{1,2,3}>0$. In the bulk we find several vertices, ranging from $+++$ to $---$. However, as it was already mentioned, not all of them can be written with the help of totally symmetric gauge potentials $\Phi_{a_1...a_s}$. With the latter we can detect only three types of vertices amplitudes: $\VV^{+s_1,+s_2,+s_3}$, $\VV^{-s_1,-s_2,-s_3}$ and $\VV^{(\pm s_1,\mp s_2,\mp s_3)}$, which are specific linear combinations of various $+++$, $++-$, $--+$ and $---$ vertices. It makes sense to form parity-odd and parity-even combinations of $+++$ and $---$. Then, the AdS/CFT dictionary we expect on general grounds is
\begin{align}\label{generaldict}
    \begin{pmatrix}
    \VV^{(\pm s_1,\mp s_2,\mp s_3)}\\
    \VV^{+s_1,+s_2,+s_3}+\VV^{-s_1,-s_2,-s_3}\\
    \VV^{+s_1,+s_2,+s_3}-\VV^{-s_1,-s_2,-s_3}
    \end{pmatrix}&= 
\left(
\begin{array}{ccc}
 x  & x  & 0 \\
 y  & -y  & 0 \\
 0  & 0  & z \\
\end{array}
\right)\begin{pmatrix}
    \langle J_{s_1} J_{s_2} J_{s_3} \rangle_{\text{F.B.}}\\
    \langle J_{s_1} J_{s_2} J_{s_3}  \rangle_{\text{F.F.}}\\
    \langle J_{s_1} J_{s_2} J_{s_3}  \rangle_{\text{odd}}
    \end{pmatrix}
\end{align}
with zeros on their obvious places because of parity and the non-zero entries can, in principle, be all different. It is a very special feature of AdS${}_4$/CFT${}_3$-correspondence that the free parameters arrange themselves into this form: the non-abelian vertex is a sum of an equal number of bosons and fermions; the abelian parity-even vertex is the difference of equal number of bosons and fermions. If we allow for arbitrary rescalings of the basis on both sides, there is still one scale-invariant number $(-yx)/(xy)=-1$. See also \cite{Maldacena:2011jn, Maldacena:2012sf,Giombi:2011ya} for earlier comments along these lines. The parity-odd/parity-even combinations of vertices are abelian whenever the spins are within triangle, as is clear from \eqref{abelian}. For the spins outside triangle, the vertex is not identically gauge-invariant with respect to the highest spin, which leads to its non-conservation. However, the non-conservation manifests itself for the parity-odd part only, where the boundary terms coming from the two parts of the vertex do not cancel each other. The inverse transform is
\begin{align}
    \begin{pmatrix}
    \langle J_{s_1} J_{s_2} J_{s_3} \rangle_{\text{F.B.}}\\
    \langle J_{s_1} J_{s_2} J_{s_3}  \rangle_{\text{F.F.}}\\
    \langle J_{s_1} J_{s_2} J_{s_3}  \rangle_{\text{odd}}
    \end{pmatrix}&= 
\left(
\begin{array}{ccc}
 \frac{1}{2 x} & \frac{1}{2 y} & 0 \\
 \frac{1}{2 x} & -\frac{1}{2 y} & 0 \\
 0 & 0 & \frac{1}{z} \\
\end{array}
\right)\begin{pmatrix}
    \VV^{(\pm s_1,\mp s_2,\mp s_3)}\\
    \VV^{+s_1,+s_2,+s_3}+\VV^{-s_1,-s_2,-s_3}\\
    \VV^{+s_1,+s_2,+s_3}-\VV^{-s_1,-s_2,-s_3}
    \end{pmatrix}\,.
\end{align}
Similarly, for one scalar leg we have two vertices $\VV^{+s_1,+s_2,0}$ and $\VV^{-s_1,-s_2,0}$, which are again specific linear combinations of various $++0$, $+-0$ and $--0$. The dictionary is
\begin{align}
    \begin{pmatrix}
    \VV^{0,+s_1,+s_2}+\VV^{0,-s_1,-s_2}\\
    \VV^{0,+s_1,+s_2}-\VV^{0,-s_1,-s_2}
    \end{pmatrix}&= 
\left(
\begin{array}{ccc}
 x'  & 0 \\
 0  & y'
\end{array}
\right)\begin{pmatrix}
    \langle J_{0} J_{s_1} J_{s_2} \rangle_{\text{X.B.}}\\
    \langle J_{0} J_{s_1} J_{s_2}  \rangle_{\text{odd}}
    \end{pmatrix}
\end{align}
where for $\Delta=1$ ($\Delta=2$) boundary conditions we have $\text{X.B.}=\text{F.B.}$ ($\text{X.B.}=\text{C.B.}$) and the odd part corresponds to $\text{C.F.}$ ($\text{F.F.}$). The odd vertex is automatically outside triangle (unless $s_1=s_2$) and is not conserved with the respect to the leg carrying the highest spin.

Since $\langle J_{0} J_{0} J_{s}  \rangle$ is unique and corresponds to $\VV^{0,0,\pm s}$ the dictionary is trivial here. There is also a subtle vertex $\VVV^{0,0,0}$ that should give zero for the critical vector model for $\Delta=2$ and $\langle J_{0} J_{0} J_{0}  \rangle_{\text{F.B.}}$ for $\Delta=1$. This is an example of an extremal correlator. It can be approached via $4+\epsilon$ as in \cite{Bekaert:2014cea}, via a careful treatment of the boundary conditions \cite{Freedman:2016yue} or by an analytic continuation in helicity \cite{Skvortsov:2018uru}.

While the complete position/momentum space dictionary is missing at the moment, one can try to make contact with the approach of \cite{Jain:2021vrv,Jain:2021gwa,Jain:2021whr}. Any correlation function can be split into homogeneous (h) and non-homogeneous (nh) parts. Let us restrict to the case where the spins satisfy the triangle inequality and ignore contact terms as we do not have those under control in position space. The h-part satisfies trivial Ward identities and can be obtained as the difference $\langle JJJ \rangle_{F.B.}-\langle JJJ \rangle_{F.F.}$ between the free boson and free fermion correlators. This is in accordance with the AdS/CFT dictionary established in  \eqref{generaldict} where  $\langle JJJ \rangle_{F.B.}-\langle JJJ \rangle_{F.F.}$ is dual to the parity even abelian vertex, which is $\VVV^{+++}+\VVV^{---}$. The parity-odd h-part is the difference $\VVV^{+++}-\VVV^{---}$. The nh-part is given by $\langle JJJ \rangle_{F.B.}+\langle JJJ \rangle_{F.F.}$ and, as is clear from \eqref{generaldict}, is dual to the non-abelian vertex.

As is clear from the AdS/CFT dictionary, the bosonization duality demands the bulk interaction to be 
\begin{align}\label{csaction}
\begin{aligned}
   &\tfrac{1}{2x}\VV^{(\pm s_1,\mp s_2,\mp s_3)}+\VV^{0,0,\pm s} +\epsilon \VVV^{0,0,0}+\\
   &+ \tfrac{1}{2y} \cos(2\theta)[\VV^{+s_1,+s_2,+s_3}+\VV^{-s_1,-s_2,-s_3}]+\tfrac{1}{z}\sin(2\theta)[\VV^{+s_1,+s_2,+s_3}-\VV^{-s_1,-s_2,-s_3}]+\\
   &\,\,\,+\tfrac{1}{y'}\cos(\theta)[\VV^{0,+s_1,+s_2}+\VV^{0,-s_1,-s_2}]+\tfrac{1}{x'}\sin(\theta)[\VV^{0,+s_1,+s_2}-\VV^{0,-s_1,-s_2}]\,.
\end{aligned}
\end{align}
The lowest derivative non-abelian interactions, e.g. those of Yang-Mills and pure gravity, are universally present with no dependence on $\theta$ (the first line). An interesting feature noted in \cite{Misuna:2017bjb} is that $\theta=\pi/4$ is somewhat special in that the highest derivative abelian interaction vanishes and the theory is maximally parity-violating (the corresponding CFT is at the mid point between its fermionic and bosonic dual formulations).

Lastly, let us stress an appealing feature of the bulk spinor-helicity formalism we employ: while it does not allow us to separate vertices into the ones having definite helicity structure in some of the cases, it is sufficient to capture the bosonization duality. In other words, it picks the right linear combinations of the vertices with same spins but different helicities to account for Chern-Simons matter theories.

%%%%%%%%%%%%%%%%%%%%%%%%%%%%%%%%%%%%%%%%%%%%%%%%%%%%%%%%%%
\section{(Spinor)-helicity and bosonization}
\label{sec:Bosonization}
%%%%%%%%%%%%%%%%%%%%%%%%%%%%%%%%%%%%%%%%%%%%%%%%%%%%%%%%%%

Let us explain briefly the idea of \cite{Skvortsov:2018uru} on how to prove the three-dimensional bosonization duality with the help of helicity decomposition. The whole duality is due to existence of a one-parameter family of correlation functions that interpolates between two theories, one is built out of fermionic matter and another one out of the bosonic matter. In particular, we will clarify the dictionary between the fine-grained helicity decomposition of the amplitudes/vertices/correlators and the one we have in the paper which is more coarse-grained but still suffices. 

We begin with the basis of cubic vertices $\VVV^{\lambda_1,\lambda_2,\lambda_3}$ that has definite helicities on external legs. It is useful to set $\Lambda=\lambda_1+\lambda_2+\lambda_3$. Now, the sectors with $\Lambda>0$ and with $\Lambda<0$ are closed and self-consistent. They belong to (anti)-Chiral higher spin gravity \cite{Metsaev:1991mt,Metsaev:1991nb,Ponomarev:2016lrm,Ponomarev:2017nrr,Skvortsov:2018jea,Skvortsov:2020wtf,Skvortsov:2022syz,Sharapov:2022faa,Metsaev:2018xip,Skvortsov:2018uru,Krasnov:2021nsq,Sharapov:2022awp}. The importance of these two sectors is in that they are rigid, i.e. all relative couplings between various vertices are completely fixed\footnote{The assumptions that guaranteed uniqueness, see \cite{Metsaev:1991mt,Metsaev:1991nb,Ponomarev:2016lrm} for more detail, can include the presence of an at least one field with spin $s>2$ together with its nontrivial self-interaction, which leads to nontrivial contribution to Ward identities, i.e. such structures are not conserved identically, featuring contact terms at coincident points. This mild assumption forces all other spins (at least even ones, and there are variations of the same statement that incorporate flavour symmetries \cite{Skvortsov:2020wtf}) to be present for consistency and fixes all relative couplings. } to be
\begin{align}\label{cubics}
    g\VVV_3&=g\sum_{\lambda_{1,2,3}}\frac{1}{\Gamma[\lambda_1+\lambda_2+\lambda_3]} \VVV^{\lambda_1,\lambda_2,\lambda_3}\,, & \bar{g}\bar\VVV_3&=\bar{g}\sum_{\lambda_{1,2,3}}\frac{1}{\Gamma[\lambda_1+\lambda_2+\lambda_3]} \VVVb^{-\lambda_1,-\lambda_2,-\lambda_3}\,,
\end{align}
where $g$ and $\bar{g}$ are two independent complex couplings. We also used $\VVVb$ with the bar in $\VVVb_3$ to stress that $\Lambda<0$ and the vertices are conjugate of those $\VVV$ with $\Lambda>0$.    
The $\Gamma$-function in the denominators restricts the sums to the (anti)-chiral sectors. Chiral and anti-chiral interactions form a complete basis \cite{Metsaev:2018xip}. 
The spectrum of (anti)-Chiral theories matches that of Chern-Simons matter theories. (Anti)-chiral cubic interactions are preceded by the same free term:\footnote{In the light-cone gauge a massless spin-$s$ field is represented by two `scalars' $\Phi_{\pm s}$, the helicity eigen states, that are complex conjugate of each other. After some rescaling by a factor of the Poincare radius $z$ the free kinetic operator is simply the flat space D'Alambertian $\square$, in particular, it is spin-independent, see \cite{Metsaev:2018xip}.}
\begin{align}
    \VVV_2&= \sum_{\lambda} \Phi_{-\lambda} \square \Phi_{+\lambda}\,.
\end{align}
However, (anti)-Chiral theories obviously lack `half of the vertices' at the cubic order. Nevertheless, (anti)-Chiral theories are useful building blocks due to their rigidity. Any candidate dual of Chern-Simons matter theories have to contain (anti)-Chiral theories as closed subsectors, which is similar to how Yang-Mills theory and Gravity have the corresponding self-dual theories as closed subsectors. Therefore, at the cubic order any theory with the same spectrum can be obtained by gluing chiral and anti-chiral pieces together. 

The same idea can be articulated directly on the CFT side \cite{Skvortsov:2018uru} where the light-cone gauge reduces a conserved tensor $J_{a_1...a_s}$ to just two `scalars' $\JJ_{\pm s}$ that represent helicity eigen-states. 
\begin{align}
    &\VVV^{\lambda_1,\lambda_2,\lambda_3}  &&\Longleftrightarrow && \langle \JJ_{\lambda_1}\JJ_{\lambda_2}\JJ_{\lambda_3}\rangle 
\end{align}
We will use the bulk notation, but everything we say can be rephrased in the language of $\langle \JJ_{\lambda_1}\JJ_{\lambda_2}\JJ_{\lambda_3}\rangle$ without any changes. 

An obviously (unique) consistent solution is to take a theory that is a direct sum of chiral and anti-chiral interactions with the same shared free term:
\begin{align}
    \VVV_2 +g\VVV_3+\bar{g}\bar\VVV_3+... \ .
\end{align}
For arbitrary complex couplings $g$ and $\bar{g}$ this theory is non-unitary, it violates CPT, as can be seen by looking at its gravitational part
\begin{align}
    \Phi_{-2} \square \Phi_{+2}+g \VVV^{+2,+2,-2}+g \VVV^{+2,+2,+2}+\bar{g} \bar\VVV^{-2,-2,+2}+\bar{g} \bar\VVV^{-2,-2,-2}+...\,,
\end{align}
where we see the two halves of the Einstein-Hilbert vertex and the two higher derivative vertices. There is a simple unitary solution: $g=\bar{g}=|g|$, which corresponds to the free/critical boson dual for $\Delta=1$ / $\Delta=2$ boundary conditions on the scalar field (indeed, it does not violate neither unitarity nor parity). However, it is too restrictive and is not the most general one. The most general CPT-invariant solution is obtained (i) by performing an additional $U(1)$ electromagnetic phase rotation $\Phi_\lambda \rightarrow \exp[i\theta \sign (\lambda)] \Phi_\lambda$ (this transformation is the most general one that does not affect the free kinetic terms); (ii) by taking $g=|g|e^{-i\theta}$ and $\bar{g}=|g|e^{+i\theta}$. As a result we find for the spin-two sector:
\begin{align}
    \Phi_{-2} \square \Phi_{+2}+|g|( \VVV^{+2,+2,-2}+\bar\VVV^{-2,-2,+2}) +|g|e^{+2i\theta} \VVV^{+2,+2,+2}+|g|e^{-2i\theta} \bar\VVV^{-2,-2,-2}+...\,,
\end{align}
where the second term is the Einstein-Hilbert vertex and the last two can be represented as 
\begin{align}
   |g|\cos(2\theta) (\VVV^{+2,+2,+2}+\VVV^{-2,-2,-2} )+i|g|\sin(2\theta) (\VVV^{+2,+2,+2}-\VVV^{-2,-2,-2} )\,.
\end{align}
This agrees with the desired \eqref{csaction}. The same transformation does the job for all spins. To operate with different helicity structures in a more efficient way we split various $\VVV^{\lambda_1,\lambda_2,\lambda_3}$ parts of $\VVV_3$ and $\bar{\VVV}_3$ into $\VVV^{+++}$, $\VVV^{++-}$, $\VVV^{+--}$, $\VVV^{++0}$, $\VVV^{+-0}$, $\VVV^{+00}$, and complex conjugates thereof with the helicities reversed. Note that whenever the structure has both $+$ and $-$ the restriction $\Lambda>0$ for $\VVV_3$ (or $\Lambda<0$ for $\bar{\VVV}_3$) has to be satisfied. Now, the same surgery as for $s=2$ leads to the action (up to higher order terms) that is dual to Chern-Simons vector models in the large-$N$ \cite{Skvortsov:2018uru}
\begin{align}\label{csdual}
    \VVV_2+\begin{aligned}
     &|g|\left(\vphantom{\bar\VVV^{00+}} e^{+2i\theta}\VVV^{+++}+e^{+i\theta}\VVV^{++0}+(\VVV^{++-}+\VVV^{+00})+e^{-i\theta}\VVV^{+-0}+e^{-2i\theta} \VVV^{+--}\right)+\\
    &|g|\left( e^{-2i\theta} \bar\VVV^{---}+e^{-i\theta}\bar\VVV^{--0}+(\bar\VVV^{--+}+\bar\VVV^{-00})+e^{+i\theta}\bar\VVV^{+-0}+e^{+2i\theta} \bar\VVV^{-++}\right)
\end{aligned}
\end{align}
The bulk coupling $|g|$ is of order ${\tilde{N}}^{-1/2}$, where $\tilde{N}$ is an effective number of degrees of freedom, and comes as an overall factor for the three-point functions, so we drop it in what follows. It is instructive to split the result according to the number of scalars in a vertex:
\begin{align*}
    s_1-0-0&: && \VVV^{+00}+\bar\VVV^{-00}\,,\\
    s_1-s_2-0&: && \cos(\theta) (\VVV^{++0}+\bar\VVV^{--0}+\VVV^{+-0}+\bar\VVV^{+-0})+i \sin(\theta) (\VVV^{++0}-\bar\VVV^{--0}-\VVV^{+-0}+\bar\VVV^{+-0})\,,\\
    s_1-s_2-s_3&: && \cos(2\theta)(\VVV^{+++}+ \VVV^{+--}+\bar\VVV^{---}+\bar\VVV^{-++}) +
    \VVV^{++-}+\\
    & &+i&\sin(2\theta)(\VVV^{+++}- \VVV^{+--}-\bar\VVV^{---}+\bar\VVV^{-++})+\bar\VVV^{--+}\,.
\end{align*}
The result for three non-zero spins can also be rewritten as
\begin{align}
\begin{aligned}
    &\cos^2(\theta)(\VVV^{+++}+ \VVV^{+--}+\bar\VVV^{---}+\bar\VVV^{-++}+\VVV^{++-}+\bar\VVV^{--+}) +\\
    &\sin^2(\theta)(-\VVV^{+++}- \VVV^{+--}-\bar\VVV^{---}-\bar\VVV^{-++}+\VVV^{++-}+\bar\VVV^{--+})+\\
    +i&\sin(2\theta)(\VVV^{+++}- \VVV^{+--}-\bar\VVV^{---}+\bar\VVV^{-++})\,.
\end{aligned}
\end{align}
This structure is in perfect agreement with the results of \cite{Maldacena:2012sf} on the restrictions imposed by the slightly-broken higher spin symmetry. In order to massage the result into a form displaying conformal structures of the parity preserving theories 
we can consider the two cases that are dual to the parity preserving theories, which correspond to $\theta=0$ 
\begin{align}\label{freeB}
    \begin{aligned}
     &( \VVV^{+++}+\VVV^{++-}+ \VVV^{+--}+\bar\VVV^{---}+\bar\VVV^{-++}+ \bar\VVV^{--+}) +\\
     + i&(\VVV^{++0}+\VVV^{+-0}+\bar\VVV^{--0}+\bar\VVV^{+-0}) + (\VVV^{+00}+\bar\VVV^{-00})\,,
    \end{aligned}
\end{align}
and to $\theta=\pi/2$
\begin{align}\label{freeF}
    \begin{aligned}
     &( -\VVV^{+++}+\VVV^{++-}- \VVV^{+--}-\bar\VVV^{---}-\bar\VVV^{-++}+ \bar\VVV^{--+}) +\\
     + &(\VVV^{++0}-\VVV^{+-0}-\bar\VVV^{--0}+\bar\VVV^{+-0}) + (\VVV^{+00}+\bar\VVV^{-00})\,.
    \end{aligned}
\end{align}
It is very enlightening to comparing these two pure cases with the general formula. For three non-zero spins \eqref{freeB} and \eqref{freeF} correspond to the free boson and free fermion, respectively. Therefore, we find (the three point functions are normalized as in \cite{Maldacena:2012sf} rather than to ${\tilde{N}}^{-1/2}$)
\begin{align*}
    \langle J_{s_1}J_{s_2}J_{s_3}\rangle&=\tilde N \left(\cos^2\theta \langle J_{s_1}J_{s_2}J_{s_3}\rangle_{\text{F.B.}}+\sin(2\theta)\langle J_{s_1}J_{s_2}J_{s_3}\rangle_{\text{odd}}+\sin^2\theta \langle J_{s_1}J_{s_2}J_{s_3}\rangle_{\text{F.F.}}\right)\,,
\end{align*}
where the free boson/fermion blocks were identified, the rest being called the odd part. 
For the configuration $s-0-0$ the structure is unique and we have 
\begin{align}
    \langle J_{s_1}J_{0}J_{0}\rangle&=\tilde N  \langle J_{s_1}J_{0}J_{0}\rangle_{\text{F.B.}}=\tilde N  \langle J_{s_1}J_{0}J_{0}\rangle_{\text{F.F.}}\,,\notag
\end{align}
where different labels correspond to different boundary conditions, $\Delta=1$ and $\Delta=2$. Lastly, for one scalar operator inserted we find
\begin{align}
    \langle J_{s_1}J_{s_2}J_{0}\rangle&=\tilde N \left(\cos\theta \langle J_{s_1}J_{s_2}J_{0}\rangle_{\text{X.B.}}+\sin\theta\langle J_{s_1}J_{s_2}J_{0}\rangle_{\text{odd}}\right)\,,
\end{align}
where `$\text{X.B.}$' corresponds to free/critical scalar while `$\text{odd}$' corresponds to critical/free fermion for $\Delta=1$ and $\Delta=2$ boundary conditions, respectively. The elementary manipulations above were omitted in \cite{Skvortsov:2018uru}.

Several comments are in order. (1) Since the cubic interaction \eqref{csdual} is the most general one that maintains unitarity, but breaks parity, it proves the three-dimensional bosonization duality at this order. The apparent simplicity of this result is deceptive since it crucially relies on two nontrivial results: (a) chiral and anti-chiral interactions form a complete basis; (b) chiral higher spin gravity has all of its coupling fixed in terms of just one effective coupling $g$. As a result, the dual of Chern-Simons vector models (or of any other theory with such a spectrum) can be obtained by gluing (anti)-chiral pieces together in a unitary way. The bulk fields are insensitive to whether the currents they are dual to are built out of bosons or fermions. Since we could fix the three-point correlation functions from unitarity and higher spin symmetry all theories with such a spectrum should belong to this one-parameter family of correlation functions. 

(2) it is clear that the helicity decomposition allows one to build the correlation functions for Chern-Simons vector models starting from any of the pure (parity preserving) theories. For example, we can take the dual of free/critical boson, perform the helicity decomposition of the interactions. All possible  $\VVV^{\lambda_1,\lambda_2,\lambda_3}$ are at our disposal now. We can glue them back into the most general three-point vertex/correlation function that has $\theta$ as a parameter. This is a very intriguing result: all information about correlation function of Chern-Simons vector models is already contained in any of the pure (parity preserving) theories, which one finds at $\theta=0$ or $\theta=\pi/2$. If a detailed helicity decomposition is not available, one still extract a lot of information about the correlation functions by taking correlators of two pure theories $\theta=0$ and $\theta=\pi/2$.

(3) it also follows from these results that one does not need the full power of the helicity decomposition to get the right interactions/correlation functions. Indeed, the complete helicity decomposition allows us to deal with all possible theories, including non-unitary ones, e.g. self-dual Yang-Mills and self-dual Gravity. This explains why our coarse-grained amplitudes can still do the job and, in particular, allows us to identify all the elementary (in terms of the helicity) constituents thereof. For example, it is clear that the `gravitational' interactions of \cite{Vasiliev:1986bq} are, in fact, the sums of the actual minimal gravitational interaction $\VVV^{+s,-s,2}+\bar\VVV^{+s,-s,-2}$ and of the higher derivative $\VVV^{+s,+s,-2}+\bar\VVV^{-s,-s,+2}$ interaction.

The idea can be pushed to higher orders, as \cite{Skvortsov:2018uru} suggests.\footnote{Higher order analysis in the bulk can be hard/impossible to do since the non-chiral vertices are known to be more non-local than usual field theory methods allow for \cite{Bekaert:2015tva,Maldacena:2015iua,Sleight:2017pcz,Ponomarev:2017qab}. While generally positive, earlier steps \cite{Giombi:2009wh,Giombi:2010vg} also revealed some puzzles already at the three-point level, see \cite{Boulanger:2015ova} for the summary, the conclusion being is that the dual of Chern-Simons vector models is not known in any form that is amenable to calculation of correlation functions. Nevertheless, for the pure (parity preserving theories) one can use the effective actions of \cite{deMelloKoch:2018ivk,Aharony:2020omh}, treating them as actions for composite operators. Then, it is possible to identify the chiral subsector for the arguments below to work. } It is important that one does not have to invoke any bulk arguments at all. The existence of Chiral higher spin gravity indicates that there is a closed subsector of Chern-Simons vector models. Then, Chern-Simons vector models can be understood as an expansion over this subsector. The general argument is that in the helicity basis one can perform a surgery on the correlation functions by splitting them into (anti)-chiral parts and the rest, performing the $U(1)$ EM-duality rotation and assembling them back into correlation functions that depend on one additional parameter \cite{Skvortsov:2018uru,Sharapov:2022awp}. The appearance of this additional parameter can be traced to the existence of the closed (anti)-Chiral subsectors, each having its own (in general complex) coupling constant. This enlarges the number of free parameters from one for the duals of free/critical boson to two (one of them $|g|\sim {\tilde{N}}^{-1/2}$ simply counts the orders in the weak field expansion) and proves the existence of a one-parameter family of correlators connecting the parity preserving theories. 

Let us illustrate this for four-point functions. Suppose we have an effective action that computes correlation functions of any of the parity preserving theories.\footnote{As it was discussed in the previous footnote, one can take an effective action for the composite operators' interpretation of \cite{deMelloKoch:2018ivk,Aharony:2020omh}. In particular, the action of \cite{deMelloKoch:2014vnt} is tightly linked to the light-cone gauge. This interpretation does not require any subtle bulk picture. } The best we can assume is that it is a collection of cubic, quartic, etc. vertices
\begin{align}
     \VVVf&=\VVVf_2+g\VVVf_3+g^2 \VVVf_4+\mathcal{O}(g^3)\,,
\end{align}
with an effective constant $g$ being of order ${\tilde{N}}^{-1/2}$. This action satisfies the consistency relations of the form\footnote{Below is the usual (perturbatively expanded) master equation, which can be reduced slightly depending on the approach. In the light-cone gauge and in flat space, the classical consistency conditions require $[H,J^{i-}]=0$, which can be decomposed with respect to the number of fields, $H=H_2+H_3+H_4+...$, $J^{i-}=J_2^{i-}+J_3^{i-}+J_4^{i-}+...$. Now, contracting $H$ and $J^{i-}$ with anti-commuting ghosts $c$, $c_i$, we can write $\VVVf=Hc +J^{i-}c_i$ and $\delta$ is the action of the free part $H_2 c+ J^{i-}_2c_i$. Similarly, if we have a gauge invariant action $S=S_2+S_3+...$ with gauge transformations $\delta^g=\delta_0^g+\delta_1^g+...$, we can construct the usual master action $\mathcal{S}$ with $\delta=(\mathcal{S}_2, \bullet)$. Since we follow \cite{Metsaev:2018xip,Skvortsov:2018uru}, which is in the light-cone gauge, Eq. \eqref{consis} should be understood as expansion of $[P^{-},J^{-1}]=0$. }
\begin{align}\label{consis}
    g&: &\delta \VVVf_3&=0\,, && g^2: && \delta \VVVf_4+ [\VVVf_3,\VVVf_3]=0\,, && ...
\end{align}
There is nothing more we can say about it, a priory. Now, the key point is that there are two closed consistent subsectors that correspond to (anti)-Chiral higher spin gravities.\footnote{Even though these theories are in AdS${}_4$, we can easily see that they define the corresponding closed subsectors of Chern-Simons vector models since Chiral theories are local and present no problem to compute the holographic correlation function.} Therefore, there is a more fine-grained structure for each of $\VVVf$:
\begin{align}
    \VVVf_2&=\VVV_2\,, & \VVVf_3&=\VVV_3+\VVVb_3\,, & \VVVf_4&=\VVV_4+\VVVb_4+\VVVh_4\,,
\end{align}
where the free terms are exactly the same for all the three theories. The cubic terms can be decomposed into the chiral $\VVV_3$ (sum of the helicities is positive) and anti-chiral $\VVVb_3$ (sum of the helicities is negative) components and this decomposition is complete. The quartic component has the (anti)-chiral parts as well as a leftover $\VVVh_4$. Now, the consistency of the (anti)-chiral theories together with the consistency of the initial one imply
\begin{align}&
\begin{aligned}
    \delta \VVV_3&=0\,,\\
    \delta \VVVb_3&=0\,,
\end{aligned} &&
\begin{aligned}
    \delta \VVV_4+ [\VVV_3,\VVV_3]=0\,,\\
    \delta \VVVb_4+ [\VVVb_3,\VVVb_3]=0\,,\\
    \delta \VVVh_4+ [\VVV_3,\VVVb_3]=0\,.
\end{aligned}
\end{align}
Now, the fact that each of the chiral theories has its own (in general complex) coupling constant, we call $g$ and $\bar{g}$, implies that the following effective action is still consistent
\begin{align}
    \VVVf_{g,\bar{g}}&= \VVV_2+ g\VVV_3 +g^2 \VVV_4 + \bar{g}\VVVb_3 +\bar{g}^2 \VVVb_4+ g\bar{g} \VVVh_4+...
\end{align}
Already at the level of three-point functions we could see that there is a unique way to get a unitary parity violating theory: $U(1)$ electromagnetic phase rotation $\Phi_\lambda \rightarrow \exp[i\theta \sign (\lambda)] \Phi_\lambda$ combined with $g=|g|e^{-i\theta}$, $\bar{g}=|g|e^{+i\theta}$. To be more specific, let us consider the four-point functions of higher spin currents (i.e. $s>0$). We have the following helicity decomposition of the quartic vertices:
\begin{align}
    \VVV_4&= \VVV_4^{++++}+\VVV_4^{+++-}+\VVV_4^{++--}+\VVV_4^{+---}\,,\\
    \VVVb_4&= \VVVb_4^{----}+\VVVb_4^{---+}+\VVVb_4^{--++}+\VVVb_4^{-+++}\,,\\
    \VVVh_4&=\VVVh_4^{++++}+\VVVh_4^{+++-}+\VVVh_4^{++--}+\VVVh_4^{+---}+\VVVh_4^{----}\,,
\end{align}
where we took into account the constraints on the sum of the helicities for the (anti)-chiral parts. The surgery results in (we drop the overall $|g|^2$-factor)
\begin{align*}
    &\cos (2 \theta ) \left(\VVVb_4^{{----}}+\VVVb_4^{{--++}}+\VVVh_4^{{---+}}+\VVVh_4^{{-+++}}+\VVV_4^{{++--}}+\VVV_4^{{++++}}\right)+\\
    &i \sin (4 \theta ) \left(\VVVb_4^{{-+++}}-\VVVh_4^{{----}}-\VVV_4^{{+---}}+\VVVh_4^{{++++}}\right)+\\
    &\cos (4 \theta ) \left(\VVVb_4^{{-+++}}+\VVVh_4^{{----}}+\VVV_4^{{+---}}+\VVVh_4^{{++++}}\right)+\\
    -&i \sin (2 \theta ) \left(\VVVb_4^{{----}}-\VVVb_4^{{--++}}+\VVVh_4^{{---+}}-\VVVh_4^{{-+++}}+\VVV_4^{{++--}}-\VVV_4^{{++++}}\right)+\\
     &\VVVb_4^{{---+}}+\VVVh_4^{{--++}}+\VVV_4^{{+++-}}
\end{align*}
In order to build the four-point function we also need to add the `exchanges'. Since the surgery does not affect the free term, the exchanges due to the cubic vertices make contributions to each of the structures, but do not produce any new type of $\theta$-dependence. The theories at $\theta=0,\pi/2$ preserve parity. To conclude, we could show that there exists a one-parameter family of correlation functions that starts from one parity-preserving theory and ends on another one. We should stress, however, that the uniqueness of the one-parameter family of CFT's does not follow immediately and has been shown up to the three-point level only \cite{Skvortsov:2018uru}.

%%%%%%%%%%%%%%%%%%%%%%%%%%%%%%%%%%%%%%%%%%%%%%%%%%%%%%%%%%
\section{Conclusions and Discussion}
\label{sec:Conclusions}
%%%%%%%%%%%%%%%%%%%%%%%%%%%%%%%%%%%%%%%%%%%%%%%%%%%%%%%%%%
In this paper we elaborated on one of the spinor-helicity formalisms that are suitable for (A)dS${}_4$ calculations. We developed a general technique to scalarize any contact interaction. This was exemplified by a number of three-point vertices/amplitudes and the corresponding three-point functions. A complete basis of interactions in this approach was used in each of the cases to capture the three-dimensional bosonization duality.

In the last part we elaborated on the light-front bootstrap suggested in \cite{Skvortsov:2018uru}. The fine-grained helicity decomposition combined with the rigidity of Chiral higher spin gravity makes it easy to fix the most general unitary but parity-violating cubic interactions and, hence, the three-point correlation functions. Remarkably, once the helicity decomposition is available one can build the three-point functions of Chern-Simons vector models out of any of the pure (parity preserving) limits thereof. This indicates that the large-$N$ Chern-Simons vector models are free theories in disguise in the sense that they contain exactly the same information but repackaged in a different way (the correlation functions do not coincide with any of the free theories, but can be built from such). 

Several approaches to prove the bosonization duality look promising. One proof \cite{Sharapov:2018kjz,Gerasimenko:2021sxj,Sharapov:2022eiy} relies on the study of the slightly-broken higher spin symmetry \cite{Maldacena:2012sf}. It has already been shown \cite{Sharapov:2020quq,Gerasimenko:2021sxj} that the slightly-broken higher spin symmetry admits unique invariants that deform those of the unbroken higher spin symmetry. The latter were computed in \cite{Colombo:2012jx,Didenko:2012tv,Didenko:2013bj, Bonezzi:2017vha} and shown to given the correct correlation functions of higher spin currents of the parity-preserving large-$N$ CFT's. What remains to be done is to actually compute the deformed invariants. The second promising approach is again based on the slightly-broken higher spin symmetry and amounts to analyzing the its constraints with many positive results already available
\cite{Li:2019twz,Kalloor:2019xjb,Turiaci:2018nua,Jain:2020puw,Jain:2021vrv,Jain:2021gwa,Jain:2021whr,Silva:2021ece}, especially the very recent \cite{Jain:2022ajd}. The third approach, which makes the proof elementary, is via Chiral higher spin gravity AdS${}_4$ \cite{Skvortsov:2018uru,Sharapov:2022awp}, whose mere existence proves that Chern-Simons vector models have two closed subsectors. This observation allows one to introduce one more coupling constant and show that there is a one-parameter family of correlators interpolating between the parity preserving theories (it does not prove yet that there cannot be other free parameters, in principle). It would be beneficial to relate all these approaches to each other. 

%%%%%%%%%%%%%%%%%%%%%%%%%%%%%%%%%%%%%%%%%%%%%%%%%%%%%%%%%%
\begin{appendix}
%%%%%%%%%%%%%%%%%%%%%%%%%%%%%%%%%%%%%%%%%%%%%%%%%%%%%%%%%%

%%%%%%%%%%%%%%%%%%%%%%%%%%%%%%%%%%%%%%%%%%%%%%%%%%%%%%%%%%
\section*{Acknowledgments}
\label{sec:Aknowledgements}
%%%%%%%%%%%%%%%%%%%%%%%%%%%%%%%%%%%%%%%%%%%%%%%%%%%%%%%%%%
We are grateful to Euihun Joung, Kirill Krasnov, Dmitry Ponomarev and Alexander Zhiboedov for useful discussions. E.S. expresses his gratitude to the Erwin Schr{\"o}dinger Institute in
Vienna for hospitality during the program ``Geometry for Higher Spin Gravity: Conformal Structures, PDEs, and Q-manifolds'' while this work was in progress. This project has received funding from the European Research Council (ERC) under the European Union’s Horizon 2020 research and innovation programme (grant agreement No 101002551). Y.Y. acknowledges the hospitality of UMONS during the course of this work and support by Fonds de la Recherche Scientifique - FNRS under Grant No. Grants No. F.4503.20 and F.4544.21. The work of Y.Y. is also supported by
\textquotedblleft the Fundamental Research Funds for the Central
Universities, NO. NS2020054\textquotedblright\ of China.
%%%%%%%%%%%%%%%%%%%%%%%%%%%%%%%%%%%%%%%%%%%%%%%%%%%%%%%%%%%%%%%%%%%%%%%
\section{ Notation and Conventions}
\label{app:notation}
%%%%%%%%%%%%%%%%%%%%%%%%%%%%%%%%%%%%%%%%%%%%%%%%%%%%%%%%%%%%%%%%%%%%%%%
We adopt the mostly plus convention for the metric $\eta_{\mm\nn} = (-+++)$, which makes the  Euclidean rotation easier to implement. 
Choosing $x^\mm=(\vec \rmx, z)$ 
to be Poincar\'e coordinates with $z$ being the radial coordinate and $\vec\rmx$ the three 
coordinates along the boundary, the AdS${}_4$-background can be described by vierbein 
$h^{\ga\gad}_\mm$ and spin-connection that splits into (anti) self-dual parts 
$\varpi_{\mm}{}^{\ga\gb}$ and ${\bar\varpi}_{\mm}{}^{\gad\gbd}$:
\begin{align}
h^{\ga\gad} &=\frac1{2z} \sigma_\mm^{\ga\gad} dx^\mm\ , 
& \varpi^{\ga\gb}& = \frac{i}{2z}\vec \sigma^{\ga\gb} \cdot d\vec x\ , &
{\bar\varpi}^{\gad\gbd}& = -\frac{i}{2z}\vec\sigma^{\ga\gb}\cdot d\vec x\ . 
\end{align}
The matrices $\sigma_\mm^{\ga\gad}$ are constant and in our convention they are given by
$\sigma_\mm^{\ga\gbd}= (\vec \sigma^{\ga\gb}, i \epsilon^{\alpha\beta})$. We have the relations\footnote{The bulk Lorentz algebra, $sl(2,\mathbb{C})$, is not manifest in the Poincar\'e coordinates. The boundary Lorentz algebra, $sl(2,\mathbb{R})$, is still manifest. A possible source of confusion is the existence of $\epsilon^{\ga\gad}$ that allows us to translate between dotted and undotted indices. As a result the $3d$ coordinates $\rmx$ can carry different types of indices $\rmx^{\ga\gb}$, $\rmx^{\ga\gbd}$ or $\rmx^{\gad\gbd}$, all representing exactly the same two-by-two matrix. }
\begin{align}
\rmx^{\ga\gb} &= \vec \sigma^{\ga\gb}\cdot  \vec \rmx\ ,  & 
{\rmx}^2 & = -\tfrac12 \rmx^{\ga\gb}\rmx_{\ga\gb}\ , &
\sigma^{\ga\gad}_\mm \sigma_{\nn\,\ga\gad} &=-2\eta_{\mm\nn}\ ,
\w2
 x^{\ga\gad} &=\rmx^{\ga\gad}+iz\epsilon^{\ga\gad}\ , &
x^2 &= \rmx^2 + z^2\ , &\rmx_{ij} &= |\rmx_i-\rmx_j|\ .
\end{align}
The inverse vierbein $h_{\ga\gad}^\mm = -z \sigma^\mm_{\ga\gad}$ obeys the relations 
\be
h^{\ga\gad}_\mm h^\nn_{\ga\gad}=\delta^\nn_\mm\ ,\qquad 
h^{\ga\gad}_\mm h^\mm_{\gb\gbd}=\delta_{\gb}^{\ga}\delta_{\gbd}^{\gad}\ , 
\ee
and the AdS${}_4$ metric tensor is given by
\bea
g_{\mm\nn}dx^\mm dx^\nn  &=& -h^{\ga\gad}_\mm h\fd{\nn\,\ga\gad}dx^\mm dx^\nn 
=\frac1{2z^2} \eta_{\mm\nn}dx^\mm dx^\nn\ .
\eea
We use the raising and lowering conventions: $X^\alpha = \epsilon^{\alpha\beta} X_\beta$ and  
$X_\alpha = X^\beta \epsilon_{\beta\alpha}$ with 
$\epsilon_{\alpha\gamma}\epsilon^{\beta\gamma} =\delta_\alpha^\beta$, 
and similar conventions for the dotted indices. We define $\epsilon_{12}=-\epsilon_{21}=1$. It is sometimes convenient to define $\nabla_{\ga\gad}$ as $\nabla=\nabla_{\ga\gad} h^{\ga\gad}_\mm \, dx^\mm$. The Lorentz covariant derivative is defined as
\begin{align}
\nabla V^{\ga\gad}  \equiv dV^{\ga\gad} -\varpi\fud{\ga}{\gb}V^{\gb\gad}
-{\bar\varpi}\fud{\gad}{\gbd}V^{\ga\gbd}\ .
\label{LorentzDerA}
\end{align}
\paragraph{Vierbein.} Background vierbein $h^{\ga\gad}$ can be used to define the basis of two-, three- and four-forms:
\begin{align}
H^{\gad\gad}&=h\fdu{\nu}{\gad}\wedge h^{\nu\gad}\,, &
H^{\ga\ga}&=h\fud{\ga}{\gnd}\wedge h^{\ga\gnd}\,,
& h\fud{\ga}{\gnd}\wedge H^{\gbd\gnd}&=\hat{h}^{\ga\gbd}\,,
\end{align}
which obey certain useful identities:
\begin{align}\label{hidentities}
\begin{aligned}
h^{\ga\gad}\wedge h^{\gb\gbd}&=\frac12  H^{\ga\gb}\epsilon^{\gad\gbd}+\frac12H^{\gad\gbd}\epsilon^{\ga\gb}\,, \qquad\qquad &   H^{\ga\ga}\wedge H^{\gad\gad}&=0\,,\\
h^{\ga\gad}\wedge H^{\gb\gb}&=-\frac23 \epsilon^{\ga\gb}\hat{h}^{\gb\gad}\,,  &
h^{\ga\gad}\wedge \bar H^{\gbd\gbd}&=+\frac23 \epsilon^{\gad\gbd} \hat{h}^{\ga\gbd}\,,\\
h^{\ga\gad}\wedge \hat{h}^{\gb\gbd}&=-\frac14 \epsilon^{\ga\gb}\epsilon^{\gad\gbd}H_{\ga\ga}\wedge H^{\ga\ga}\,,\\
H_{\ga\ga}\wedge H^{\ga\ga}&=-H_{\gad\gad}\wedge H^{\gad\gad}\,, & H_{\ga\ga}\wedge H^{\ga\ga}&=-h_{\ga\gad}\wedge \hat{h}^{\ga\gad}\,,\\
H^{\ga\gb}\wedge H^{\gc\gd}&=\frac16(\epsilon^{\ga\gc}\epsilon^{\gb\gd}+\epsilon^{\gb\gc}\epsilon^{\ga\gd}) H_{\nu\nu}\wedge H^{\nu\nu}\,.
\end{aligned}
\end{align}
The volume form is defined to be $\vol=H_{\nu\nu}\wedge H^{\nu\nu}$. In the Poincare coordinates we have
\begin{align}
H^{\ga\gb}&=\frac{1}{4z^2}(d\rmx\fud{\ga}{\gnd}\wedge d\rmx^{\gb\gnd}+2i d\rmx^{\ga\gb}\wedge dz)\,,\\
\vol&=\frac{i}{2z^4} d^3\rmx dz\,,\\
\hat{h}^{\ga\gad}\big|_{dz=0}&=\frac{1}{8z^3}\epsilon^{\ga\gad} d^3\rmx\,.
\end{align}
We also define $|\vol|=i/2$ as a factor of difference between the simplest AdS${}_4$ invariant volume $d^3\rmx dz/z^4$ and $\vol$. The $i$ will go away in Euclidian signature. We will ignore this $1/2$ in the main text.

%%%%%%%%%%%%%%%%%%%%%%%%%%%%%%%%%%%%%%%%%%%%%%%%%%%%%%%%%%%%%%%%%%%%%%%
\section{Identities, relations}
\label{app:identities}
%%%%%%%%%%%%%%%%%%%%%%%%%%%%%%%%%%%%%%%%%%%%%%%%%%%%%%%%%%%%%%%%%%%%%%%
\paragraph{Identities between $\boldsymbol{P,Q,S}$.} As explained in \cite{Giombi:2011rz}, three points plus three polarization spinors make $15$ free parameters in total. The conformal group has dimension $10$ and, hence, there should be only $5$ truly free parameters. On the other hand, we have nine structures $P,Q,S$. The even power of any odd structure is even, which is manifested by \cite{Giombi:2011rz}
\begin{align}
\begin{aligned}
S_1^2+Q_2 Q_3-P_{1}^2\equiv0\,,\\
S_2^2+Q_1 Q_3-P_{2}^2\equiv0\,,\\
S_3^2+Q_1 Q_2-P_{3}^2\equiv0\,,\\
P_{2} Q_2+P_{3} P_{1}+S_1 S_3\equiv0\,,\\
P_{3} Q_3+P_{1} P_{2}+S_1 S_2\equiv0\,,\\
P_{1} Q_1+P_{3} P_{2}+S_2 S_3\equiv0\,.
\end{aligned}
\end{align}
There is also another relation \cite{Giombi:2011rz}
\begin{align}
    P_{3}^2 Q_3+P_{1}^2 Q_1+P_{2}^2 Q_2+2 P_{1} P_{2} P_{3}-Q_1 Q_2 Q_3\equiv 0\,.
\end{align}
All these relations, which are consequences of the Fierz (or Schouten) identities, bring a lot of ambiguity in how any correlation function can be written. 

\paragraph{Differential relations.} The relations below prove that $K$, $\tanV_{\ga\gad}$, $\xi^\ga$ and $\brxi^\gad$ form a closed set under all reasonable differential and algebraic manipulations. We begin with 
\begin{align}
\nabla \tanV^{\ga\gad}=h^{\ga\gad}+\tanV\fud{\ga}{\gdd}h^{\gd\gdd}\tanV\fdu{\gd}{\gad}=0\ ,
\end{align}
and the parallel transported spinors obey
\begin{align}
\nabla\xi^\ga-\tanV\fud{\ga}{\gdd} \xi_\gd h^{\gd\gdd}=0\ ,\qquad
\nabla\bar\xi^{\gad}-\tanV\fdu{\gd}{\gad} \bar\xi_{\gdd} h^{\gd\gdd}=0\ .
\label{AppbtobD}
\end{align}
In practice it is useful to rewrite the Lorentz-covariant derivatives with all 
indices being explicit:
\begin{align}
\nabla_{\ga\gad}K &= K \tanV_{\ga\gad}\ ,& \nabla_{\ga\gad} \xi_\beta &=\tanV_{\beta\gad}\xi_\ga\  , &
\nabla_{\ga\gad} \brxi_{\dot\beta} &=\tanV_{\ga\dot\beta}\bar\xi_\gad\ , &
\nabla_{\ga\gad} \tanV_{\gb\gbd} &=2\epsilon_{\ga\gb}\epsilon_{\gad\gbd}+\tanV_{\ga\gad}\tanV_{\gb\gbd}\ .
\end{align}
As a consequence of the differential constraints above one also finds
\begin{align}
(\square-4) K &=0\ , & (\square-6)\tanV^{\ga\gad} &=0\ ,
& (\square-4)\xi^{\ga}&=0\ , & (\square-4)\brxi^{\gad}&=0\ .
\end{align}
$K$ is a boundary-to-bulk propagator for a scalar field that is dual to a $\Delta=1$ operator on the boundary. The first Eq. determines our normalization of the cosmological constant: $m^2=\Lambda \Delta(\Delta-d)=2\Lambda=-4$.

%%%%%%%%%%%%%%%%%%%%%%%%%%%%%%%%%%%%%%%%%%%%%%%%%%%%%%%%%%%%%%%%%%%%%%%
\paragraph{Algebraic Identities.} 
%%%%%%%%%%%%%%%%%%%%%%%%%%%%%%%%%%%%%%%%%%%%%%%%%%%%%%%%%%%%%%%%%%%%%%%
The algebraic relations begin with
\be
\tanV^{\ga\gad}\tanV\fud{\gb}{\gad}=\epsilon^{\ga\gb}\ ,\qquad
\tanV^{\ga\gad}\tanV\fdu{\ga}{\gbd}=\epsilon^{\gad\gbd}\ .
\label{FisSpTwo}
\ee
and continue as
\begin{align}
\tanV\fud{\ga}{\gdd}\Pib^{\gdd\gb} &= i\Pi^{\ga\gb}\ , 
& \tanV\fdu{\gc}{\gad}\Pi^{\gc\gb} &=-i\Pib^{\gad\gb}\ , 
& \Pi\fud{\ga}{\gc}\Pib^{\gad\gc}=-iKF^{\ga\gad}\ ,
& \\
\xi^{\ga} &=\tanV\fud{\ga}{\gad}\bar{\xi}^\gad\ , 
& \bar{\xi}^{\gad} &={\xi}^\ga \tanV\fdu{\ga}{\gad}\ .
\end{align}
The parallel-transport bi-spinors $\Pi^{\ga\gb}$ and $\Pib^{\gad\gb}$ satisfy identities 
similar to \eqref{FisSpTwo}:
\begin{align}
\Pi^{\ga\gb}\Pi\fud{\gc}{\gb} &=K \epsilon^{\ga\gc}\ , &
\Pib^{\gad\gb}\Pib\fud{\dot\gc}{\gb} &=K \epsilon^{\gad\dot\gc}\ ,
\\
\Pi^{\gb\ga}\Pi\fdu{\gb}{\gc} &=K \epsilon^{\ga\gc}\ , &
\Pib^{\gbd\ga}\Pib\fdu{\gbd}{\gc} &=K \epsilon^{\ga\gc}\ .
\end{align}
There are also useful identities involving the one-form 
$h^{\alpha\dot\alpha} =dx^\mm h_\mm^{\alpha\dot\alpha}$:
\begin{align}
&(\tanV\cdot h) \tanV^{\ga\gad}+h^{\ga\gad}=\tanV\fud{\ga}{\gbd}h^{\gb\gbd}\tanV\fdu{\gb}{\gad}\ ,
\\
&(\tanV\cdot h) \xi^\alpha +(\tanV\fud{\ga}{\gdd}h\fdu{\gb}{\gdd}-\tanV_{\gb\gdd}h^{\ga\gdd})\xi^\gb=0\ ,
\end{align}
which are due to the Fierz identities.

%%%%%%%%%%%%%%%%%%%%%%%%%%%%%%%%%%%%%%%%%%%%%%%%%%%%%%%%%%%%%%%%%%%%%%%
\paragraph{Inversion Map.}
%%%%%%%%%%%%%%%%%%%%%%%%%%%%%%%%%%%%%%%%%%%%%%%%%%%%%%%%%%%%%%%%%%%%%%%
In addition to \eqref{prime} we need the action of inversion $R$ on the bulk coordinates
\begin{align}
    Rx^{\ga\gad}&=\frac{x^{\ga\gad}}{x^2}=\frac{\rmx^{\ga\gad}+iz\epsilon^{\ga\gad}}{\rmx^2+z^2}\ .
\end{align}
Together with \eq{prime} we can now derive the action of $R$ on 
$K$, $\tanV$, $\xi$ and $\bar{\xi}$:
\begin{align}
\begin{aligned}
K(R(\rmx,z);R\rmx_i)&= \rmx_i^2 K(\rmx,z;\rmx_i)\ ,
\w2
\Pi^{\ga\beta}(R(\rmx,z);R(\rmx_i))&
=-J\fud{\ga}{\gdd}\bar{\Pi}\fud{\gdd}{\delta}(\rmx,z;\rmx_i)\rmx_i^{\delta\beta}\ ,
\w2
\xi^{\ga}(R(\rmx,z);R(\rmx_i,\eta_i))&=+iJ\fud{\ga}{\gdd}\bar{\xi}^{\gdd}(\rmx,z;\rmx_i,\eta_i)\ ,
\w2
\bar{\xi}^{\gad}(R(\rmx,z);R(\rmx_i,\eta_i))&=-iJ\fdu{\gc}{\gad}{\xi}^{\gc}(\rmx,z;\rmx_i,\eta_i)\ ,
\w2
\tanV^{\ga\gad}(R(\rmx,z);R(\rmx_i))&=J\fud{\ga}{\gbd} J\fdu{\gb}{\gad}\tanV^{\gb\gbd}(\rmx,z;\rmx_i)\ ,
\end{aligned}
\end{align}
where we defined $J^{\ga\gad}$ as
\begin{align}
J^{\ga\gad}&=\frac{x^{\ga\gad}}{\sqrt{x^2}}=\frac{\rmx^{\ga\gad}
+iz\epsilon^{\ga\gad}}{\sqrt{\rmx^2+z^2}}\,, && J\fud{\ga}{\gdd}J^{\gb\gdd}=-\epsilon^{\ga\gb}\ .
\end{align}
%

%%%%%%%%%%%%%%%%%%%%%%%%%%%%%%%%%%%%%%%%%%%%%%%%%%%%%%%%%%%%%%%%%%%%%%%
\section{Three-point functions}
\label{app:algorithm}
%%%%%%%%%%%%%%%%%%%%%%%%%%%%%%%%%%%%%%%%%%%%%%%%%%%%%%%%%%%%%%%%%%%%%%%
We confine the algorithm of computing the correlation functions in the Appendix for the reason of its being conceptually simple but a bit technical. However, as far as we know the general algorithm of how to scalarize any AdS-integrand has not been given yet. We concentrate on the AdS${}_4$ case, where the spinorial language can be used, but the procedure can be used for any $d$. Continuing the discussion of section \ref{sec:holocoralg} we apply the inversion map $R$ both to the boundary and bulk data, see Appendix \ref{app:identities} for the action of $R$. The resulting dictionary reads:
{\allowdisplaybreaks
\besubeqs
\begin{align}
\frac{d^3\rmx dz}{z^4}&\rightarrow \frac{d^3\rmx dz}{z^4}\ ,
\\
K_1&\rightarrow z\ ,
\\
K_{2,3}&\rightarrow \rmx_{2,3}^2 K_{2,3}\ ,
\\
(\xi_1 \tanV_2\bar{\xi}_1) &\rightarrow 2 z K_2 [\eta_1 (\rmx-\rmx_2)\eta_1]=-2 z K_2T_{11}^2\ ,
\\
(\xi_1 \tanV_3\bar{\xi}_1) &\rightarrow 2 z K_3 [\eta_1 (\rmx-\rmx_3)\eta_1]=-2 z K_3T_{11}^3\ ,
\\
(\xi_2 \tanV_1\bar{\xi}_2) &\rightarrow 2 K_2^2 [\eta_2 (\rmx-\rmx_2)\eta_2]=-2 K_2^2T_{22}^2\ ,
\\
(\xi_3 \tanV_1\bar{\xi}_3) &\rightarrow 2 K_3^2[\eta_3 (\rmx-\rmx_3)\eta_3]=-2 K_3^2T_{33}^3\ ,
\\
\left(\xi _{2}\tanV_{3}\bar{\xi}_{2}\right) &\rightarrow 2K_{2}K_{3}
\left( 
\frac{(\rmx_{23})^{2}}{z}K_{2}\left[ \eta _{2}\left( \rmx-\rmx_{2}\right)
\eta _{2}\right] 
+\left[ \eta _{2}\left( \rmx_{2}-\rmx_{3}\right) \eta _{2}\right] 
\right)  \ ,
\\
(\xi_1\xi_2)+(\bar\xi_1\bar\xi_2)&\rightarrow 2 z K_2 (\eta_1\eta_2)\ ,
\\
(\xi_1\xi_3)+(\bar\xi_1\bar\xi_3)&\rightarrow 2 z K_3 (\eta_1\eta_3)\ ,
\\
(\xi_1\xi_2)-(\bar\xi_1\bar\xi_2)&\rightarrow -2 i K_2 [\eta_1 (\rmx-\rmx_2)\eta_2]=2 i K_2 T_{12}^2\ ,\\
(\xi_1\xi_3)-(\bar\xi_1\bar\xi_3)&\rightarrow -2 i K_3 [\eta_1 (\rmx-\rmx_3)\eta_3]=2 i K_3 T_{13}^3\ ,\\
\left( \xi _{2}\xi _{3}\right) +\left( \bar{\xi}_{2}\bar{\xi}_{3}\right) 
&\rightarrow 2K_{2}K_{3}\left[ \eta _{2}\rmx_{23}\eta _{3}\right]  \ ,
\\
\left( \xi _{2}\xi _{3}\right) -\left( \bar{\xi}_{2}\bar{\xi}_{3}\right) 
&\rightarrow 2K_{2}K_{3}\eta _{2\alpha }\eta _{3\beta }
\left[ 
\frac{1}{iz}\left( \rmx-\rmx_{2}\right) ^{\alpha \gamma }\left( \rmx-\rmx_{3}\right) _{\gamma }{}^{\beta }-iz\epsilon ^{\alpha \beta }
\right] \ ,
\end{align}
\esubeqs}
\noindent where we defined
\begin{align}\label{bulktij}
T_{ij}^l&=-[\eta^i_\ga (\rmx-\rmx_l)^{\ga\gb}\eta^j_\gb]\,.
\end{align}
We presented the complete list of the structures that are relevant for any bosonic theory in the bulk. In theories with fermions one can also find few more structures, e.g. $\left(\xi _{2}\tanV_{1}\bar{\xi}_{3}\right)$, see Appendix \ref{app:exotic}, which can be treated in the same way.   

Now it is obvious how to rewrite the integrand in terms of simple differential operators acting on a scalar integrand of type 
\be
\int \frac{d^3\rmx dz}{z^4} z^a (K_2)^b (K_3)^c= (\rmx_{23})^{a-b-c} I_{a,b,c}\ ,
\label{mi}
\ee
where
\be
I_{a,b,c}=\frac{\pi ^{3/2} \Gamma \left(\frac{1}{2} (a+b-c )\right) \Gamma 
\left(\frac{1}{2} (a-b+c )\right) \Gamma \left(\frac{1}{2} (-a+b+c )\right) 
\Gamma \left(\frac{1}{2} (a+b+c -3)\right)}{2 \Gamma (a) \Gamma (b) \Gamma (c )}\ .
\ee
To scalarize the integrand we define $O^l_{ij}\equiv (\eta_i \pl_l \eta_j)\equiv \eta^i_\ga \pl_{\rmx_l}^{\ga\gb} \eta^j_\gb$. The main property of $O^k_{ij}$ is that it generates the corresponding $T^l_{ij}$ from $K_l$, $l=2,3$:
\begin{align}
O_{ij}^l f(K_l)&= \frac{(K_l)^2}{z} \frac{\pl}{\pl K_l} f(K_l) \,T_{ij}^l \ .   
\end{align}
However, after several applications of the formula operators $O^l_{ij}$ can hit $T^l_{i'j'}$ as well, which is manifested by
\begin{align}
    O^{a}_{bc}\, T^{a'}_{b'c'}&= \delta^{aa'} \tfrac12 [(\eta_b \eta_{b'})(\eta_c \eta_{c'})+(\eta_b \eta_{c'})(\eta_c \eta_{b'})]\,.
\end{align}
As is clear from the last relation, the extra terms, which are various purely boundary quantities $(\eta_a \eta_b)$, are of lower order in $T$'s. Therefore, one can subtract them by adding a polynomial in $O$'s of a lower order as well, and so on. In the bulk, we have the following structures:
\begin{align}
    T^2_{11}\,, T^2_{22}\,, T^2_{12}\,, T^3_{11}\,, T^3_{33}\,, T^3_{13} \,, \qquad
    Y\equiv\eta _{2\alpha }\eta _{3\beta } (\rmx-\rmx_{2})^{\alpha \gamma }(\rmx-\rmx_{3}) _{\gamma }{}^{\beta }\,.
\end{align}
All $T$'s can be scalarized by the procedure just described. In principle, one can introduce a new second order operator to generate $Y$ from $K_2K_3$, but there is a simpler way. Let us introduce two auxiliary polarization spinors $\eta_4$ and $\eta_5$ and apply the identity
\begin{align}\label{Ystr}
  \pl_{\eta_4}^\nu \pl^{\eta_5}_\nu  (T^2_{24} T^3_{35})&=Y\,.
\end{align}
As a result, $Y$ gets replaced with $(T^2_{24} T^3_{35})$, the latter can be treated as all the other $T$'s. At the end one has to perform contractions of $\eta_{4,5}$. Now, the integrand is reduced to 
\begin{align}\label{sumi}
    V_3&= \sum_\aI g_\aI[(\eta_i\eta_j), (\eta_i\rmx_{23}\eta_j), O^{k}_{ij}]\, ( z^{a_\aI} K_2^{b_\aI} K_3^{c_\aI}) \ ,
\end{align}
where $(\eta_i\rmx_{23}\eta_j)\equiv \eta^i_\ga (\rmx_2-\rmx_3)^{\ga\gb} \eta^j_\gb$ and the sum contains a finite number of terms. It is important that $O$'s act on $x_{2,3}$ in $K_2$, $K_3$ and not on $(\eta_i\rmx_{23}\eta_j)$. In Appendix \ref{app:scalarizeagain} we outline an even simpler procedure to scalarize integrands.

\paragraph{Back to $\boldsymbol{P,Q,S}$.}
The integral can now be done term by term in \eqref{sumi}, after which the operators $O$ should be applied. Technically, it is convenient to use the same notation $T_{ij}$ for the corresponding structures on the boundary $T_{ij}=[\eta_i \rmx_{23} \eta_j]$ since they arise under the action of $O_{ij}^2$ on $\rmx_{23}$ resulting from the integral
\begin{align}
    O_{ij}^2 f(\rmx_{23})&= \frac{1}{\rmx_{23}} T_{ij}^2 \frac{\pl}{\pl \rmx_{23}} f(\rmx_{23}) \ .
\end{align}
Note that $O^3_{ij}$ gives exactly the same but with the minus sign. Afterwards, one should apply the following dictionary to recover the conformally invariant structures:
{\allowdisplaybreaks
\besubeqs\label{cfti}
\begin{align}
\begin{aligned}
&\begin{aligned}
\rmx_{2,3}&\rightarrow \frac{1}{\rmx_{2,3}}\ ,
\\
\rmx_{23}&\rightarrow \frac{\rmx_{23}}{\rmx_2 \rmx_3}\ ,
\end{aligned}&&\qquad\qquad
\begin{aligned}
P_{12}&\rightarrow (\eta_1\eta_2)\ ,
\\
P_{23} &\rightarrow + \left[\eta _{2} \rmx_{23} \eta _{3} \right]\frac{1}{\mathrm{x}_{23}^{2}} \ ,
\\
P_{31}&\rightarrow (\eta_3\eta_1)\ ,
\end{aligned}\\
&\begin{aligned}
Q_1&\rightarrow -[\eta_1 \rmx_{23}\eta_1]\ ,
\\
Q_2&\rightarrow +\left[\eta_2 {\rmx_{23}}{}\eta_2\right]\frac{1}{\rmx_{23}^2}\ ,
\\
Q_{3} &\rightarrow +\left[ \eta _{3}x_{23}\eta _{3}\right] \frac{1}{\mathrm{x}_{23}^{2}} \ ,
\end{aligned}&&\qquad\qquad
\begin{aligned}
S_{1} &\rightarrow \frac{\left( \eta _{2}\eta _{3}\right) }{\rmx_{23}}\ ,
\\
S_{2} &\rightarrow +\left[ \eta _{1}\rmx_{23}\eta _{3}\right] \frac{1}{\rmx_{23}}\ ,\\
S_3&\rightarrow +\left[\eta_1 {\rmx_{23}}\eta_2\right]\frac{1}{\rmx_{23}}\ .
\end{aligned}
\end{aligned}
\end{align}
\esubeqs}\noindent
%

%%%%%%%%%%%%%%%%%%%%%%%%%%%%%%%%%%%%%%%%%%%%%%%%%%%%%%%%%%%%%%%%%%%%%%%
\section{Beyond three-point}
\label{app:npoint}
%%%%%%%%%%%%%%%%%%%%%%%%%%%%%%%%%%%%%%%%%%%%%%%%%%%%%%%%%%%%%%%%%%%%%%%
The scalarization algorithm we described in Appendix \ref{app:algorithm} works for any number of legs (obviously, there is one leg, say $x_1$, on which we play the inversion map, all the others make no difference). Therefore, we can compute a contribution to an $n$-point function from any contact interaction vertex in the bulk. By `computing' we mean to express it in terms of the scalar bulk star integral. For $n=4$ this is known as $D$-function \cite{DHoker:1999kzh}, which denotes (up to some self-evident prefactor) the integral with $\Delta_i$ boundary-to-bulk propagators as $D_{\Delta_1,\Delta_2,\Delta_3,\Delta_4}$. For example \cite{Arutyunov:2000ku}, 
\begin{align}
    D_{1,1,1,1}(z,\bar{z})&= \frac{1}{z-\brz }\left( 2 \mathrm{Li}_2(\tfrac{1}{\brz})- 2 \mathrm{Li}_2(\tfrac{1}{z})+ \log (z\brz) \log \frac{\brz(z-1)}{z(\brz-1)}\right)\,,
\end{align}
where $z$, $\brz$ are related to the two cross-ratios, $z \brz =u$, $(1-z)(1-\brz)=v$. It is clear that any function of $\xi_{ij}^+$, where
\begin{align}
    \xi_{ij}^\pm&= \tfrac12[ (\xi_i \xi_j) \pm (\brxi_i \brxi_j)]\,,
\end{align}
reduces to the overall factor with $\xi_{ij}^+$ replaced by $P_{ij}$. For other cases, scalarization results in derivatives of $D$-function. One can use various relations that $D$ obeys as well as its series expansion, \cite{DHoker:1999kzh,Arutyunov:2000ku}. For example, the Seagul vertex $\phi^\dag A_m A^m \phi$ in the scalar QED leads to
\begin{align}
    \int \vol\phi^\dag A_{\ga\gad} A^{\ga\gad} \phi=\int \vol K_1 K_2 K_3 K_4 \xisc{2}{3} \xiscb{2}{3}\,.
\end{align}
The quartic vertex of Yang-Mills theory looks very simple in the spinor helicity formalism:
\begin{align}
   \Tr \int \vol A_{\ga\gnd}A\fdu{\ga}{\gnd} A\fud{\ga}{\gmd} A^{\ga\gmd}=\int \vol K_1 K_2 K_3 K_4 (\brxi_2 \brxi_1)(\brxi_4 \brxi_3)(\xi_3 \xi_1)(\xi_4 \xi_2) +\text{perm.} (12)(34)
\end{align}
Likewise, for gravity we find (for the holomorphic part)
\begin{align*}
    \int &Q_{\ga\ga} Q^{\ga\ga}= \tfrac{1}{48}\int \vol  [\xisc{2}{1}+\xiscb{2}{1}][\xisc{4}{3}+\xiscb{4}{3}][\xisc{4}{2}\xisc{3}{1}+\xisc{4}{1}\xisc{3}{2}]\times \\
    &\times \big\{ \xiscb{2}{1}\xiscb{4}{3}[\xisc{4}{2}\xisc{3}{1}+\xisc{4}{1}\xisc{3}{2}]+\xisc{2}{1}\xisc{4}{3}[\xiscb{4}{2}\xiscb{3}{1}+\xiscb{4}{1}\xiscb{3}{2}]\big\}\,,
\end{align*}
where $Q_{\ga\ga}=e\fud{\ga}{\gdd}\wedge e^{\ga\gdd}+\omega\fud{\ga}{\gb}\wedge \omega^{\ga\gb}$. The problem of star integrals becomes simpler in momentum space and in the light-cone gauge. For example, for $d=3$ a spin-$s$ field is represented by a pair of complex scalars $\Phi_{\pm s}(p,z)$, which are conjugate to each other, see e.g. \cite{Metsaev:2018xip}. The simplest case is $\Delta=1$, for which the boundary-to-bulk propagator is $|p|^{-1}\exp{(-|p|z])}$ (after some rescaling of $\Phi$ by a factor of $z$). The bulk integral is then trivial. 

%%%%%%%%%%%%%%%%%%%%%%%%%%%%%%%%%%%%%%%%%%%%%%%%%%%%%%%%%%%%%%%%%%%%%%%
\section{Exotic structures}
\label{app:exotic}
%%%%%%%%%%%%%%%%%%%%%%%%%%%%%%%%%%%%%%%%%%%%%%%%%%%%%%%%%%%%%%%%%%%%%%%
Due to integration by parts and Fierz identities there can be more than one way to present any given interaction. In the main text we tabulated the structures that are sufficient for any bosonic theory. In theories with fermionic fields one can find two more
\begin{align*}
&\left(\xi _{2}\tanV_{1}\bar{\xi}_{3}\right) \rightarrow K_{2}K_{3}
\left(
\frac{1}{iz}\eta _{2\alpha }\left( \rmx-\rmx_{2}\right) ^{\alpha \gamma }\left( \rmx-\rmx_{3}\right)_{\gamma }{}^{\beta }\eta _{3\beta }
+\left[ \eta _{2}\left( \rmx-\rmx_{2}\right) \eta _{3}\right] 
+\left[ \eta _{2}\left( \rmx-\rmx_{3}\right) \eta _{3}\right] 
-iz\left( \eta_{2}\eta _{3}\right)
\right) 
\\
&\left(\xi _{1}\tanV_{3}\bar{\xi}_{2}\right) \rightarrow K_{2}K_{3}\eta _{1\alpha }\eta_{2\beta }
\left\{ 
\frac{1}{iz}\left( \rmx-\rmx_{3}\right)^{2}\left( \rmx-\rmx_{2}\right)^{\alpha \beta }
-\left( \rmx-\rmx_{3}\right) ^{\alpha \gamma }\left[ \left( \rmx-\rmx_{2}\right) 
-\left( \rmx_{2}-\rmx_{3}\right) \right] _{\gamma}{}^{\beta }
\right. \nonumber\\ &\qquad \qquad \qquad \quad \left. 
-iz\left[ \left( \rmx-\rmx_{3}\right) +\left( \rmx_{2}-\rmx_{3}\right) \right] ^{\alpha \beta }
-z^{2}\epsilon ^{\alpha\beta }
\right\} \\
&\left( \xi _{1}^{\alpha }\tanV_{2\alpha }{}^{\dot{\gamma}}\right) \left( \xi
_{1}^{\beta }\tanV_{3\beta \dot{\gamma}}\right)  \rightarrow \eta _{1\alpha
}\eta _{1\beta }\left[ -4z^{2}K_{2}K_{3}\left( \mathrm{x}_{2}-\mathrm{x}%
_{3}\right) ^{\alpha \beta }-4izK_{2}K_{3}\left( \mathrm{x}-\mathrm{x}%
_{2}\right) ^{\alpha \gamma }\left( \mathrm{x}-\mathrm{x}_{3}\right)
_{\gamma }{}^{\beta }\right.   \notag \\
&\ \ \ \ \ \ \ \ \ \ \ \ \ \ \ 
\left. -2zK_{2}\left( \mathrm{x}-\mathrm{x}_{2}\right)
^{\alpha \beta }+2zK_{3}\left( \mathrm{x}-\mathrm{x}_{3}\right) ^{\alpha
\beta }\right] 
\end{align*}
and the last one can sometimes appear, but it can be simplified via integration by parts. All of these structures can be taken into account by using the same techniques as in the main part.

%%%%%%%%%%%%%%%%%%%%%%%%%%%%%%%%%%%%%%%%%%%%%%%%%%%%%%%%%%%%%%%%%%%%%%%
\section{Scalarize once again}
\label{app:scalarizeagain}
%%%%%%%%%%%%%%%%%%%%%%%%%%%%%%%%%%%%%%%%%%%%%%%%%%%%%%%%%%%%%%%%%%%%%%%
For any number of point the bulk integrand can be understood as a function of $T_{ij}^l=-[\eta^i_\ga (\rmx-\rmx_l)^{\ga\gb}\eta^j_\gb]$, $\eta_i \rmx_{kl} \eta_j$, $\eta_i\eta_j$, $z$ and $s_l=z/K_l$:
\begin{align}
   \int \frac{d^3\rmx dz}{z^4} F(\eta_i\eta_j, \eta_i \rmx_{kl} \eta_j, T_{ij}^l, s_l, z)\,.
\end{align}
It is understood that $\eta_i$ may include some auxiliary polarization spinors as to factorize $Y$ structures \eqref{Ystr}. All $\eta_i \rmx_{kl} \eta_j$ and $\eta_i\eta_j$ immediately factor out and we omit (but do not forget about) them in what follows. We want to trade all $T$'s for $O^l_{ij}\equiv (\eta_i \pl_l \eta_j)\equiv \eta^i_\ga \pl_{\rmx_l}^{\ga\gb} \eta^j_\gb$. Let us introduce label $(l,A)$ instead of $l,ij$ on $T_{ij}^l$ and $O_{ij}^l$. The starting point for scalarization is 
\begin{align}\label{applyingO}
    O_{A,l} \,f(s_i, T_{B,j})&= -T_{A,l} \frac{\pl}{\pl s_l} f +C^{l}_{A|B} \frac{\pl}{\pl T_{B,l}} f\,,
\end{align}
where `structure constants' $C^{l}_{A|B}$ depend on $\eta_i\eta_j$ according to
\begin{align}
    O^{a}_{bc}\, T^{a'}_{b'c'}&= \delta^{aa'} \tfrac12 [(\eta_b \eta_{b'})(\eta_c \eta_{c'})+(\eta_b \eta_{c'})(\eta_c \eta_{b'})]\,.
\end{align}
Therefore, $C_{A|B}=C_{B|A}$. Also note that different points $\rmx_l$ do not `see each other' since $O$'s commute for different $l$. Therefore, we drop label $l$ as well and scalarize with respect to each point separately. It is useful to consider 
\begin{align}
    \prod_A \exp[ t_A O_A] f(s)&= f\left(s-\sum_A t_A T_A -\tfrac12 \sum_{A,B} C_{A|B} t_A t_B\right)\,,
\end{align}
which can be computed with the help of $e^A e^B =e^{A+B+\tfrac12 C}$ where $C=[A,B]$ and commutes with $A$, $B$. Indeed, $O_A$ behaves like a translation operator in $s$, $T_A$. Now, we can extract the l.h.s of
\begin{align}
     &s^a \prod_A (T_A)^{b_A} && \longrightarrow && (s-\sum_A t_A T_A)^{a+\sum_C b_C}\Big|_{\prod_M t_N^{b_N}} 
\end{align}
as the coefficient of the r.h.s. The latter can be represented as
\begin{align*}
    \prod_A \exp[ t_A O_A] \exp[\tfrac12 \sum_{A,B} C_{A|B} t_A t_B \pl_s] s^{a+\sum_C b_C}&=\prod_A \exp[ t_A O_A](s+\tfrac12 \sum_{A,B} C_{A|B} t_A t_B)^{a+\sum_C b_C} \ .
\end{align*}
The last expression can be Taylor expanded as to isolate the coefficient of ${\prod_M t_N^{b_N}}$. For example, for just one variable $T$ (omitting index $A$) we have
\begin{align*}
    (s-t T)^{a+b}\Big|_{t^b}&= \frac{\Gamma[a+2b]}{b!\Gamma[a+b]}(-)^b s^a T^b\,,\\
    (s-t T)^{a+b}&=\exp[ t O] \exp[\tfrac12 Ct^2 \pl_s] s^{a+b}=\exp[ t O] (s+\tfrac12 C t^2)^{a+b}=\\
    &=\sum_n t^{n}\sum_j \frac{\Gamma[a+b+j]}{2^j(n-2j)! j! \Gamma[a+b]} C^j O^{n-2j} s^{a+b-j}\,,
\end{align*}
which allows us to derive
\begin{align}
    s^a T^b&= \sum_j \frac{\Gamma[a+b+j]}{2^j(b-2j)! j! \Gamma[a+b]} C^j O^{b-2j} s^{a+b-j}\,.
\end{align}
We can also extract the scalarized integrand $\tilde{f}(O,s)$ for any function $f(s,T)$ as
\begin{align}
    \tilde{f}(O,s)&= \mathrm{res}_t\, \frac{1}{t} e^{tO}f\left(s+\tfrac12 C t^2,\tfrac{1}{t}(s+\tfrac12 C t^2)\right)\,.
\end{align}
For three-point correlators the bulk integral gives $\rmx_{23}^\nu$ with various $\nu$'s and applying $O$'s follows the same formula as in the bulk \eqref{applyingO} for $O^2$ and produces $(-)$ of \eqref{applyingO} for $O^3$. For more points on the boundary the action of $O^l_{ij}$ on each $\rmx_{li}$ is captured by the same \eqref{applyingO}:
\begin{align}\label{applyingOA}
    O_{A,l} \,f(\rmx_{li}^2, t_{B,lj})&= -t_{A,li} \frac{\pl}{\pl \rmx_{li}^2} f +C^{l}_{A|B} \frac{\pl}{\pl t_{B,lj}} f\,,
\end{align}
where $t^B_{lj}\equiv \eta_n \rmx_{lj}\eta_m$ for $B=(mn)$.

\end{appendix}

\footnotesize
\providecommand{\href}[2]{#2}\begingroup\raggedright\endgroup

\end{document}